\documentclass[twocolumn]{aastex61} 
\usepackage{csvsimple}
\usepackage{multirow,bigdelim}
\usepackage{amsmath}

\newcommand{\rearth}{R$_\oplus$}

\newcommand{\kms}{\mbox{km s$^{-1}$}}

\usepackage{colortbl}
\definecolor{tablegray}{rgb}{0.89, 0.89, 0.89}

\begin{document}

\title{Lunar Exploration as a Probe of Ancient Venus}
\correspondingauthor{Samuel H. C. Cabot}

\email{sam.cabot@yale.edu}

\author{Samuel H. C. Cabot}
\affil{Yale University, 52 Hillhouse, New Haven, CT 06511, USA}

\author{Gregory Laughlin}
\affil{Yale University, 52 Hillhouse, New Haven, CT 06511, USA}

\begin{abstract}

An ancient Venusian rock could constrain that {planet's} history, and reveal the past existence of oceans. Such samples may persist on the Moon, which lacks an atmosphere and significant geological activity. We demonstrate that if Venus' atmosphere was at any point thin and similar to Earth's, then asteroid impacts transferred {potentially detectable} amounts of Venusian surface material to the Lunar regolith. Venus experiences an enhanced flux relative to Earth of asteroid collisions that eject lightly-shocked ($\lesssim 40$ GPa) surface material. Initial launch conditions {plus} close-encounters and resonances with Venus evolve {ejecta trajectories} into Earth-crossing orbits. Using {analytic models for crater ejecta and} \textit{N}-body simulations, we find more than $0.07\%$ of the ejecta lands on the Moon. {The Lunar regolith will contain up to 0.2 ppm Venusian material if Venus lost its water in the last 3.5 Gyr. If water was lost more than 4 Gyr ago, 0.3 ppm of the deep megaregolith is of Venusian origin.} About half of collisions between ejecta and the Moon occur at $\lesssim6$ \kms, which hydrodynamical simulations have indicated is sufficient to avoid significant shock alteration. Therefore, recovery and isotopic analyses of Venusian surface samples would determine with high confidence both whether and when Venus harbored liquid oceans and/or a lower-mass atmosphere. {Tests on} brecciated clasts in existing Lunar {samples from} Apollo missions may provide an immediate resolution. Alternatively, regolith characterization by upcoming Lunar missions may provide answers to these {fundamental} questions {surrounding Venus'} evolution.

\end{abstract}


\section{Introduction} \label{sec:intro}

We focus on an important unsolved Solar System problem: did Venus at one point harbor Earth-like conditions? Did it host liquid water, and if so, when were its oceans depleted? 

Modern-day Venus, with its $\sim90$ bar CO$_2$-dominated atmosphere and its $\sim735$ K surface \citep{Marov1978}, contrasts sharply with Earth. Yet Venus and Earth have similar bulk masses and radii, and they drew from the same region of the Solar Nebula. Their primordial compositions were likely similar \citep{Walker1975, Zharkov1983}. High atmospheric D/H ratios suggest that Venus initially had abundant water \citep{Donahue1982, Hamano2013} either as oceans or as atmospheric vapor. A runaway greenhouse process was later established \citep{Walker1975}, and the Venusian water was photo-dissociated and subsequently depleted via hydrodynamic escape of hydrogen \citep{Watson1981, Donahue1997, Kasting1988}. The remaining oxygen may have reacted with magma oceans \citep{Kasting1988}, or escaped non-thermally \citep{Shizgal1996}; solar winds can remove the dissociated ions, as observed by ESA's {\it Venus Express} \citep{Delva2008, Persson2018}. 

In light of its interesting and complex past, theoretical and experimental efforts have been made to better constrain the ancient Venusian surface and atmosphere. While the {atmospheric} D/H ratio provides strong evidence for an initially large abundance of water, this quantity can be affected by planetesimal impacts, formation conditions, and outgassing \citep{Lammer2009}. Initial condensation into liquid oceans may have been possible due to the Sun's reduced luminosity following formation \citep{Kasting1988}. The timeline for the greenhouse runaway has some constraints. Radiogenic argon measurements favor early and/or quick outgassing \citep{Kaula1999}, placing most of the hydrogen loss prior to $\sim$~3.5 Gya. A review of Venus probes, chemical and isotopic analyses, and radar mapping is given by \citet{Fegley2003}, and more recent missions are discussed by \citet{Glaze2014}. 

In short, there is near-complete uncertainty regarding the actual evolutionary path that Venus took. As \citet{Lammer2009} notes, the true history of the planet remains elusive in the absence of a surface or subsurface sample; and acquiring such samples pose significant challenges given the present conditions on Venus. Even if sample return was feasible, the current surface is probably not very telling of Venus' distant past. Observations indicate craters have a typical age of only 0.5 Gyr \citep{Schaber1992}, possibly due to a catastrophic resurfacing event. The mantle now lies below a thick (stagnant) lithosphere \citep{Solomatov1996, Moresi1998}. 

We investigate an alternative route to Venusian sample recovery: identification of ejecta on the Moon originating from a cataclysmic impact on Venus. An often-cited example on Earth is the Chicxulub impact \citep{Schulte2010}, largely held responsible for the Cretaceous–Paleogene (K-Pg, also known as the K-T) mass extinction event approximately 66 Myr ago. The $\sim$10 km wide asteroid impact left a 200 km diameter crater, triggered a tsunami, and ejected over $10^{16}$ kg of material into a vapor plume \citep{Kring2002, Schulte2010, Gulick2019}. Speculated consequences include widespread firestorms, shock heating of the atmosphere, global reduction in sunlight, and acidic rain \citep{Lewis1982, Pope1994, Robertson2013, Gulick2019}. Importantly, \citet{Kring2002} estimate that 12$\%$ of material reached escape velocity, assisted by isotropic plume expansion in the upper atmosphere and Earth's rotation. These particles have potential to reach other planets in the Solar System \citep{Reyes-Ruiz2012}. 

As an order-of-magnitude estimate for the frequency of events involving at least $\sim$10 km diameter impactors, \citet{Mileikowsky2000} find 11.0 impacts per Gyr, with $3.2\times10^9$ total fragments ejected from Earth (even more fragments considering those heated above $100^\circ$C). Roughly $\sim 0.1\%$ reach Mars. While K-T magnitude events are predicted to occur {on Earth} every $\sim10^8$ years \citep{Chapman1994}, millions of particles are ejected by smaller impacts, which occur at $\sim 10^6$ year intervals. Ejection of lightly-shocked rock {(i.e. subject to $\lesssim$ 40 GPa of pressure)} requires impact velocities of at least twice the escape velocity of the planet \citep{Melosh1984}. Comet impact speeds often exceed this value, whereas asteroids rarely hit Earth with sufficient speed. Venus on the other hand experiences significantly faster asteroid impacts since it lies closer to the Sun.  Ejecta age and origin may be pinpointed by their isotope and mineral content \citep{Wood1982, Bogard1983b}. One remarkable case is Martian meteorite ALH84001 \citep{McKay1996, Keprta2001, Golden2001, Martel2012, Mathew2001, Halevy2011}. Similarly, if Mercurian or Venusian meteorites were to arrive on Earth, their origin could potentially be inferred from their composition \citep[][and references therein]{Righter2006}. 

Delivery from Venus has not been explored thoroughly in the literature, in part because it has a high escape velocity relative to Mercury and Mars, and lies deeper in the Sun's gravitational potential compared to Earth. Its thick present-day atmosphere is, however, the largest deterrent. Ejecta would have more readily escaped Venus during its posited Earth-like phase. This period may have lasted for only a short while after formation, or it may have extended to as recently as $\sim$0.7 Gya \citep{Way2016}. 

While Earth resurfaces on timescales of $\sim$500 million years \citep{Sobolev2011}, the Moon lacks geological activity and offers promise to preserve {records of ancient impacts near its surface} \citep[e.g.][]{Joy2012, Joy2016}. This prospect is considered by \citet{Armstrong2002} in the context of Earth fragments during {Late Heavy Bombardment} (LHB).  While \citet{Armstrong2002} also mention the possibility of recovering Venusian ejecta, their analysis was based on a short $N$-body simulation that generated a small statistical sample. The renewed interest in Lunar exploration and ability to leverage significant computational resources invites a re-evaulation of this prospect.

Recent years have seen a world-wide resurgence in Lunar exploration initiatives \citep{Lawrence2017, Taylor2018}. This trend is underscored by the United States Space Policy Directive 1, which states near-term intent to return humans to the Moon. The active NASA Artemis\footnote{www.nasa.gov/specials/artemis/} program specifies a goal to establish a Lunar base by 2024. Collaborators include ESA, JAXA and CSA. {Relevant missions led by other countries include China's {\it Change'e} probe and lander series, Russia's Luna-Glob orbiters and landers, and India's {\it Chandrayaan-3} lander.} These efforts {reflect} global interests to utilize Lunar resources and establish a permanent presence on the Moon (both {crewed} and robotic), and they prime the pump for further Solar System exploration. In contrast to the Space Race of the Cold War era, Lunar exploration is drawing substantial independent participation by private-sector companies \citep{Voosen2018}. Contractors for NASA's Commercial Lunar Payload Services (CLPS) program include SpaceX, Blue Origin, Astrobotic Technology and eleven others for launch, transport, lander, and rover services. These companies also receive funding from smaller bodies or private individuals for independent space missions. While these missions are largely focused on human expansion into outer space, many involve, at the very least, secondary science goals encompassing geological characterization, {and will} greatly assist {in} obtaining a more detailed Lunar geological history \citep{Wasserburg1987}.

We show through a combination of impact mechanics and numerical simulations that asteroid collisions with Venus placed detectable quantities of ejecta on the Moon over the duration of any extended period that Venus had a thin atmosphere. Our goal is to specifically quantify the amount of Venusian material on the Moon as a function of the onset date for Venus' runaway greenhouse. We also seek to quantify the ratio of Venusian to Earth material in the Lunar regolith, and to estimate the uncertainties for these bottom-line quantities. 

We adhere to the following organization. We use secular theory to model the transfer of ejecta from the orbit of Venus to the orbit of Earth in \S\ref{sec:secular}. This provides a base-line analytic model that can be compared  with the full $N$-body simulations which we describe in \S\ref{sec:methods} and \S\ref{sec:res}. These simulations resolve the Earth-Moon system within a full Solar System model, and they adopt initial conditions that typify the immediate outcomes of various high-energy impacts on Venus. In \S\ref{sec:ej}, we draw on Monte-Carlo simulations that sample the distributions of plausible impactor velocities, sizes, and events to obtain a density (and associated uncertainty) of Venusian material on the Moon (expressed as ppm in the Lunar regolith) as a function of water-loss time. We discuss further considerations and specific routes to sample recovery in \S\ref{sec:disc}. In summary, our calculations point to realistic near-term prospects for obtaining ancient Venusian surface samples that will clarify the planet's evolutionary path.

\section{Secular Evolution of Ejecta} 
\label{sec:secular}

Our analysis suggests that the Solar System's architecture is remarkably well suited for producing transport of minimally altered Venusian rocks to the Moon. It is thus important to understand why the process is so efficient.
Before studying full numerical simulations of the transfer of ejecta between Solar System bodies, preliminary insight can be gained from the  predictions from the classical second-order secular theory. 

After carrying out an expansion of the gravitational disturbing function to first order in mass, and second order in eccentricity and inclination, \citet{Laplace1784} famously determined that
\begin{equation}
\frac{d}{dt}
        \begin{bmatrix}
           z_{1} \\
           \vdots \\
           z_{k} \\
           \zeta_{1} \\
           \vdots \\
           \zeta_{k}
         \end{bmatrix}
        =
\sqrt{-1} 
\begin{bmatrix}
A_k \:\:\: 0_k\\
0_k \:\:\: B_k
\end{bmatrix}
        \begin{bmatrix}
           z_{1} \\
           \vdots \\
           z_{k} \\
           \zeta_{1} \\
           \vdots \\
           \zeta_{k}
         \end{bmatrix}\,,
\end{equation}
where $z = e\exp{\sqrt{-1}\varpi}$, and $\zeta = \sin{i/2}\exp{\sqrt{-1}\Omega}$, for orbital eccentricity $e$, inclination $i$, longitude of periastron $\varpi$, and longitude of ascending node $\Omega$ {(see \citet{Laskar1996} for additional details)}. The block matrices $A_k$ and $B_k$ depend on the semi-major axes of the $k$ planets, which are assumed constant. Zero-block matrices of appropriate shape are denoted $0_k$. This approximation neglects terms in the disturbing function expansion which contain mean longitudes, as they are rapidly varying and average to zero in absence of mean-motion resonances. At this order, the time-evolution of the eccentricities and inclinations are decoupled:
\begin{equation}
    z_i = \sum_{j=1}^{k} \alpha_{ij}e^{\sqrt{-1}g_j t}\, ,
\end{equation}
\begin{equation}
    \zeta_i = \sum_{j=1}^{k} \beta_{ij}e^{\sqrt{-1}f_j t} \, ,
\end{equation}
with eigenfrequencies $f_j$, $g_j$, and eigenvectors $\{\alpha_{ij}\}$, $\{\beta_{ij}\}$, where $i, j$ take on values between $1$ and $k=8$ corresponding to the Solar System planets. 

\citet{Brouwer1950} published a modified version of this theory that incorporates 10 inclination eigenfrequencies, where the extra two approximately account for the effect of the $5$:$2$ near-commensurability between Jupiter and Saturn. Their model provides an accurate accounting of planetary motions on time scales ranging into the millions of years, while consisting merely of a compact table of values for the eigenfrequencies and eigenvectors. Within this framework, the secular evolution of a massless test particle is also deterministic, and is set by its initial (and constant) semi-major axis, $a_0$.

\begin{figure}
    \centering
    \includegraphics[width=0.45\textwidth]{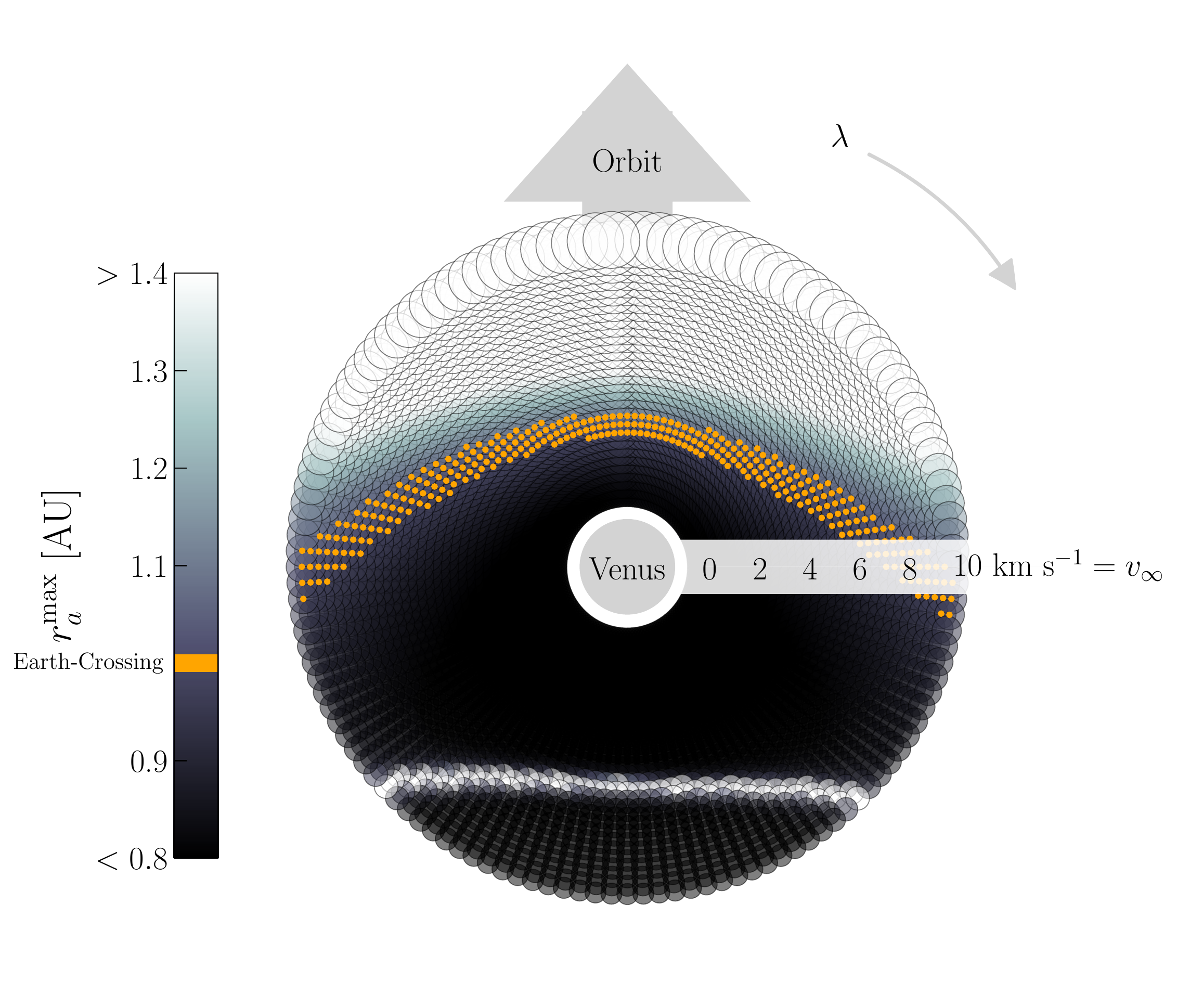}
    \caption{Maximum apocenter $(r_a^{\rm max})$ of particles launched from Venus throughout the course of 10 Myr of secular evolution. Particles which reach $r_a \gtrsim 1$ AU are able to cross Earth's orbit, and potentially collide with Earth or the Moon. Particles are launched within the orbital plane of Venus, with longitudes ($\lambda$) spanning the circumference of the planet. Circles denote individual particles and are arranged according to their initial longitude on Venus. Their distance from the center is proportional to their residual velocity, $0 \leq v_{\infty} \leq 10$ \kms. Also, the size of each circle is proportional to the initial semi-major axis of the particle, which range $0.48 < a_0 < 1.91$ AU. {No particles launched from the trailing side of Venus can reach Earth for plausible $v_{\infty}$ except for those in the lightly-shaded band}. The band in the low portion arises from secular resonance, involving two eccentricity eigenfrequencies and particles with $a_0 \approx 0.54$ AU. Particles launched from the leading hemisphere that attain $r_a^{\rm max} \approx 1$ AU are highlighted in orange.}
    \label{fig:secular}
\end{figure}

Adopting the theory, we can examine the secular evolution of particles launched from zero-latitude at Venus' Hill radius:
\begin{equation}
R_H \approx a(1-e)\sqrt[3]{\frac{m}{3M_{\odot}}}\, .
\end{equation}
Particles are given initial residual velocities
\begin{equation}
    v_{\infty} = \sqrt{v^2 - v_{\rm esc}^2}\,,
\end{equation}
between 0.0 and 10.0 \kms\ ($v_{\rm esc}$ is the escape velocity of the planet).
For each particle, the secular evolution is sampled throughout a 10 Myr history. We focus on the particle's apocenter, defined as:
\begin{equation}
    r_a(t) = a_0(1+e(t))\, .
\end{equation}
This quantity must exceed $\gtrsim 1$ AU in order to cross Earth's orbit. The maximum apocenter $(r_a^{\rm max})$ for various launch conditions is shown in Figure~\ref{fig:secular}.
As expected, particles launched closer to the leading face of Venus and with higher launch velocity attain larger orbits. Of the 3968 different combinations of initial conditions sampled, 1013 particles cross Earth's orbit from their initial orbital parameters. An additional 201 attain Earth-crossing orbits over the 10 Myr time period due to secular evolution. A region of initial conditions corresponding to $a_0 \approx 0.54$ achieves resonance with the $g_3=17.32832$ $^{\prime\prime}$yr$^{-1}$ and $g_4=18.00233$ $^{\prime\prime}$yr$^{-1}$ eigenfrequencies given by \citet{Brouwer1950}, and are forced to high eccentricity. The results show that a considerable fraction of ejecta, of order $\sim30\%$, are likely to attain Earth-crossing orbits for physically plausible initial conditions, and underscore Venus' basic propensity for delivering material to the Earth. 

Of course, the actual orbital evolution of ejecta, which is profoundly affected by energy-changing close encounters, is markedly more complex, as is the distribution of launch velocities. The need to refine the transfer rate estimates to incorporate these effects thus motivates full numerical simulations.

\section{Numerical Integrations} \label{sec:methods}

An extensive literature covers the investigation of the transfer of {rocks} between Earth and the rest of Solar System. Foregoing studies include some priors on transfer rates and provide precedents for certain computational methods. While the majority of meteorites that arrive on Earth are of asteroidal origin, studies have investigated fragments that might travel from other planets, including Mercury and Venus \citep{Gladman1996, Gladman2009, Melosh1993} and even to other moons, such as Ganymede \citep{Alvarellos2002}. Massive, high-velocity impacts are responsible for ejecting planetary material at speeds exceeding the escape velocity. The relevant mechanics are discussed at length by \citet{Melosh1989}. Simulations show that most of this material is ejected from the Solar System by Jupiter, transported inward, or recaptured by the origin planet \citep{Melosh1993, Worth2013}, but also suggest that a significant fraction of Venusian meteorites can reach Earth \citep{Melosh1993}. The reverse flow, in which Earth ejecta reaches other planets, is also well-studied \citep{Gladman2005, Reyes-Ruiz2012}, particularly in the context of the Chicxulub Impact \citep{Kring2002, Gulick2019}. Many of these studies seek to understand ejecta as a mechanism of life transport throughout the Solar System \citep[e.g.][]{Tobias1974, Melosh1988, Mileikowsky2000}. Studies also constrain ejecta velocity \citep{Melosh1989}, size \citep{Grady1980}, and number \citep{Mileikowsky2000}. The Moon has been recognized as a means of archiving information from early Solar System \citep{Crawford2008, Halim2019}; this has brought attention to the flux of Earth material onto its surface \citep{Armstrong2002, Armstrong2010}, as well as the survivability of such ejecta \citep{Burchell2010, Crawford2014, Joy2016, Halim2020}.

\subsection{Ejection Theory}
\label{sec:theory}

\begin{figure*}
    \centering
    \includegraphics[width=\textwidth]{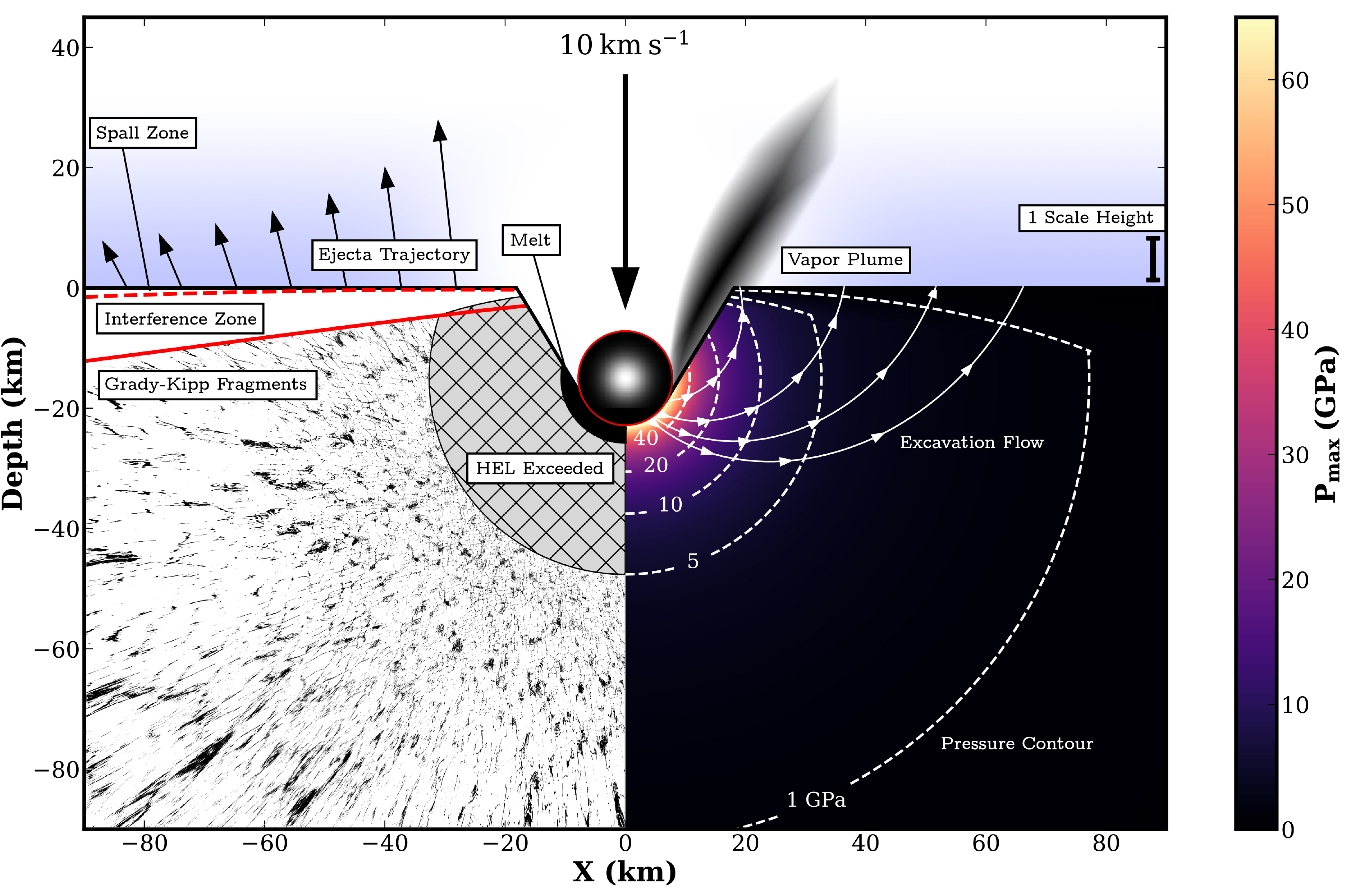}
    \caption{Schematic of a vertical impact at the surface of a terrestrial body. This example depicts the resulting processes from a 10 \kms\ projectile, including: spallation of near-surface material, which may reach the escape velocity; the interference zone which reduces the maximum pressure ($P_{\rm max}$) experienced by the rock; ejection of a vapor plume; rock undergoing plastic deformation at pressures exceeding the Hugoniot Elastic Limit (HEL); melting in the immediate vicinity of the impact; and Grady-Kipp fragmentation of rock deeper into the target. At large depths, rock fragments into bigger pieces due to reduced maximum pressure. Excavation flows clear out the region which becomes the crater. The contours of constant pressure are hemispherical below the impact, but sharply turn inwards at the interference zone. {The x-axis corresponds to horizontal distance from the impact site}. Adapted from \citep{Melosh1984}.}
    \label{fig:imp}
\end{figure*}

We approximate the quantity and properties of ejected matter using the impact cratering prescriptions of \citet{Melosh1984, Melosh1989}, in a similar manner as \citet{Mileikowsky2000}. In a hypervelocity impact, a projectile strikes a target and delivers pressure waves which exceed the material sound speed. The impact may be studied in three stages: contact and compression, excavation, and modification \citep{Melosh1989} (reviewed in Figure~\ref{fig:imp}). Contact and compression involves transfer of the projectile's kinetic energy into the ground in the form of shock waves, generating pressures up to hundreds of GPa. A shock wave also travels through the projectile to its back surface, and reflects as a rarefaction wave which destroys the projectile and continues into the target. The excavation stage involves the expansion of a massive plume of vaporized material, and decay of the shock wave into an elastic wave as it propagates into the target. Material continues to flow behind the shock wave and clears out a cavity. The modification phase comprises landslides and collapses which alter the crater topography. Contact and compression is most relevant for our analysis; however each stage is discussed in detail by \citet{Melosh1989}.

The contact and compression stage starts when {an (idealized)} spherical projectile of diameter $D$ and speed $V_i$ makes first contact with a flat target. We consider a vertical impact, but the following approximations extend to oblique impacts \citep{Melosh1984}. The projectile enters the target during the {\it rise time}:
\begin{equation}
    \tau \simeq \frac{D}{2V_i}\,,
\end{equation}
{which, for the impactor size ranges of interest in this paper (hundreds of meters to tens of kilometers), and typical impact velocities at Venus ($\sim$15 to 40 \kms), lasts from $\sim10^{-3}$ to $10^0$ seconds}. The contact delivers a shock wave pulse into the target. In the near-surface region, the longitudinal stress is approximated by the elastic Hugoniot equation:
\begin{equation}
    P_L \simeq \rho_t u_t c_L\,,
\end{equation}
with uncompressed target density $\rho_t$, target material velocity $u_t$, and P-wave speed $c_L$. Positive and negative stresses correspond to compression and tension respectively. Both $P_L$ and $u_t$ increase during the rise time, and decrease during the {\it decay time}:
\begin{equation}
    \beta\tau \simeq \frac{D}{c_L}\,,
\end{equation}
or the interval during which the reflected rarefaction wave traverses the projectile. The factor $\beta$ is greater than unity. Varying degrees of shock cause rock to undergo plastic deformation ($\gtrsim$5 GPa) or even melt ($\gtrsim$40 GPa, depending on the material). Shock waves propagate in a hemispherical shape from the points of contact. Since the projectile of density $\rho_p$ is buried within the target, the center of the spherical wavefront is located at an {\it equivalent center} of depth:
\begin{equation}
    d_{{\rm eq}} \simeq D\Big(\frac{\rho_p}{\rho_t}\Big)^{1/2}\,.
\end{equation}
At a distance $r$ from the equivalent center, $u_t$ is given by a decreasing power-law with exponent 1.87 \citep{Perret1975}. 
\citet{Gault1963} show that if the projectile and target have identical composition, then $u_t = V_i/2$ at the impact site. \citet{Melosh1984} thus give the particle velocity as a function of distance:
\begin{equation}
    u_t(r) \simeq C_V \frac{V_i}{2}\Big(\frac{D}{2r}\Big)^{1.87}\,,
\end{equation}
where $C_V$ is a {\it coupling constant} of order unity, which can account for density differences between projectile and target.

Under the free surface boundary condition, the compressional shock wave must be met with a tensional rarefaction wave of equal strength and opposite sign reflecting off the target surface. In the shallow {\it interference zone}, reflected rarefaction waves encounter material before the rise time is over. That is, the material first experiences the shock wave, which alone would compress the material to a pressure $P_{\rm free}$. Before the material reaches its maximum pressure from the shock, it experiences the rarefaction wave, and by interference, the pressure is reduced to $P_{\rm max}$. A hyperbola separates the interference zone from the lower {\it field-free zone} \citep{Melosh1989}:
\begin{equation}
    z_P = \frac{c_L\tau}{2}\sqrt{1 + \frac{s^2}{d^2_{\rm eq} + (c_L\tau/2)^2}}\,.
\end{equation}
Here, $s$ is the horizontal distance from the impact site. This interference and reduction in maximum pressure may explain why recovered meteorites from Mars are only lightly shocked \citep{Melosh1984}. 
At a depth $z$, the interfering waves produce tensional pressures:
\begin{equation}
    P_{\rm min} \simeq \frac{-2d}{r_0}\frac{z}{\beta c_L \tau}\Big(1 - 1.87\frac{\beta c_L \tau}{r_0}\Big)P(r_0),
\end{equation}
where $r_0 = \sqrt{d_{\rm eq}^2 + s^2}$, and $P(r) = \rho_t u_t(r) c_L$. If this pressure exceeds the tensile strength $T$, a layer of rock breaks off and accelerates from the pressure gradient at the surface in a process called {\it spalling}. 
\citet{Melosh1989} shows that the spall velocity is primarily vertical, and given by:
\begin{equation}
    v_{\rm ej} \simeq 2u_t(r)\Big(1 + \Big(\frac{s}{d_{\rm eq}}\Big)^2\Big)^{-1/2}\,.
\end{equation}
The velocity is upper-bounded by half the projectile velocity $V_i$ \citep{Melosh1984, Melosh1985}. The spalled material is ejected faster than any other rock affected by the impact, but is only lightly shocked compared to deeper material. Spall sheets only account for a few percent of the total ejecta mass, the rest mostly attributed to excavation flow, jetting and the vapor plume. The spall has thickness:
\begin{equation}
    l_s \simeq \frac{T}{\rho_{t} c_L v_{\rm ej}}d_{\rm eq}\,,
\end{equation}
where $T$ is the target material dynamic tensile strength \citep{Melosh1984}. Integrating this equation \citep{Melosh1985, Armstrong2002} gives the spall mass $m_{\rm ej}$ ejected at velocities of at least $v_{\rm ej}$, as a fraction of impactor mass $m_{i}$:
\begin{equation}
\label{eqn:ratio}
   \frac{m_{\rm ej}}{m_i} = \frac{0.75P_{\rm max}}{\rho_t c_L V_i}\Big[\Big(\frac{V_i}{2v_{\rm ej}}\Big)^{5/3} - 1 \Big]\,.
\end{equation}
\citet{Gladman2005} convert this equation into the fraction of total mass which reaches the escape velocity:
\begin{equation}
    F(v_{\rm esc} < v < v_{\rm ej}) = \frac{1 - (v_{\rm ej}/v_{\rm esc})^{-5/3}}{1 - (V_i/2v_{\rm esc})^{-5/3}}\,.
\label{eqn:frac}
\end{equation}
Following high-speed impacts, the spall can contain sufficient energy to break into Grady-Kipp fragments with mean size:
\begin{equation}
    L_{GK} = \frac{T}{\rho_t v_{\rm ej}^{2/3}V_i^{4/3}}D\,,
\end{equation}
\citep{Grady1980, Melosh1984, Melosh1985}. The cumulative distribution of fragment diameters depends on Weibull's constant $m$ for the target material and the maximum fragment size $L_{\rm max} = L_{GK}(m + 3)/2$ \citep{Mileikowsky2000}, and is discussed thoroughly by \citet{Melosh1992}.
In \S\ref{sec:ej}, we follow the analysis of \citet{Mileikowsky2000} for a single impact. By assuming impactor mass and speed and ejected rock material properties, we estimate the amount of mass which reaches the escape velocity of Venus. From the distribution of fragment sizes, we can approximate the number of ejected fragments, and subsequently apply the transfer rate calculated in \S\ref{sec:res}. 

We note that \citet{Chyba1994} provides an estimate of the total mass of material ejected from an impact. However, we restrict our analysis to spall material described by \citet{Melosh1985}. Heavily shocked rock may lose information about Venus' water history, or not be identifiable as Venusian in origin. In particular, zircon grains undergo severe shock-metamorphism at pressures $\gtrsim$50 GPa and decomposition past $\gtrsim$90 GPa \citep{Kusaba1985, Wittmann2006}.

\subsection{Choice of Integrator}
\label{sec:int}

Numerical simulations of ejecta orbits are common in the literature \citep{Gladman1996, Gladman2005, Gladman2009}. Recent studies \citep{Reyes-Ruiz2012, Worth2013} use the hybrid symplectic $N$-body integrator, \verb|MERCURY| \citep{Chambers1999}, which preserves energy. 
A particle enters a close encounter with a planet when it is within a multiple of the planet's Hill radius 
(e.g. ejecta studied by \citet{Reyes-Ruiz2012} enter a close-encounter when they are within three Hill radii of a planet or moon). These hybrid schemes are much faster than pure adaptive timestep schemes for integrations involving many particles and close encounters.

To model the trajectories of impact ejecta to other bodies in the Solar System, we use the $N$-body code \verb|REBOUND|\footnote{Simulations in this paper made use of the REBOUND code which is freely available at http://github.com/hannorein/rebound.} \citep{Rein2012}, and the hybrid symplectic integrator \verb|MERCURIUS| \citep{Rein2019}. \verb|MERCURIUS| is similar to \verb|MERCURY|, but {uses different criteria for identifying and resolving close-encounters}. Keplerian orbits are integrated symplectically with \verb|WHFast| \citep{Rein2015b}. Close-encounters use the 15$^{\rm th}$-order adaptive timestepping integrator \verb|IAS15| \citep{Rein2015}. {The initial \verb|WHFast| timestep is 1 hour, which remains the maximum allowed timestep for the first year of evolution. After then, the \verb|WHFast| timestep is increased to 1 day} \citep{Worth2013}. The minimum adaptive timestep is $\sim5$ seconds, which prevents the simulation from hanging on close-encounters. We integrate the trajectories in batches of 20 ejecta particles per simulation. A 10,000-particle simulation is thus distributed across 500 cores. Ejecta are classified as `test particles' which cannot collide with each other and have no gravitational influence over any other particle, including planets and the Sun. When an ejecta particle is within the radius of a planet or the Sun, we record the collision and remove the particle from the simulation. Impact ejecta which attain distances of $>100$ AU are classified as ejected from the Solar System and removed from the simulation.

\subsection{Overview of Simulations}
\label{sec:init}

We perform three sets of Simulations, which are described as follows:

\begin{itemize}
    \item Simulation 1 (calibrations): We use initial conditions similar to those in previous studies \citep{Gladman2009, Reyes-Ruiz2012, Worth2013} {in order to verify our simulation architecture}. We distribute 10,000 particles at the planet Hill radius, drawing latitude and longitude from independent, uniform distributions. Particles are given a random velocity ($v_{\rm ej}$) at $1-2$ times the escape velocity ($v_{\rm esc}$) in the radial outward direction, plus the heliocentric orbital motion of the origin planet. The simulation is integrated for 10 Myr, {during which we record collisions in which particles come within the physical radius of a planet or the Sun}. We perform simulations for particles ejected from Venus, Earth and Mars and compare to transfer rates found by \citet{Worth2013}. 
    
    \item Simulation 2 (impact and spallation): 
    We include the Moon as an active particle in the simulation {at its present-day separation from Earth}. Since the Moon is within Earth's Hill Sphere, the simulation adjusts to a smaller timestep in order to resolve its orbit. Hence we only integrate for 1 Myr. We discuss implications of this decision in \S\ref{sec:res}. {We sample velocities from a distribution corresponding to ejection from the surface by a vertical impact, and repeat for three impact velocities: (a) $25$ \kms; (b) $30$ \kms; and (c) $60$ \kms.  Our choices for impact velocities correspond to: (a) an asteroid impact slightly faster than the minimum required for spallation; (b) our nominal projectile (\S\ref{sec:ej}); and (c) a comet impact (neglecting corrections for target/projectile density). The simulation is performed for ejecta from Earth and Venus, with particles starting at $1.001\times$ the planet radii. This allows a small fraction from Earth to collide with the Moon on their way out. For Venus, ejecta velocities range between $v_{\rm esc}$ and half the projectile impact velocity (these choices are discussed in \S\ref{sec:ej} and based on Equation~\ref{eqn:frac})}. For Earth, the lowest launch speed is the minimum necessary to reach the Moon.

    \item Simulation 3 (enlarged Earth particle): 
    We repeat Simulation 2 (b), except we remove the Moon, artificially increase the radius of the Earth particle {to the Moon's current orbital separation}, and start particles from the planet's Hill radius. We integrate for 10 Myr, and obtain a larger sample from which we can derive typical speeds at which ejecta enter the Earth-Moon system. We record the velocities of colliding particles.

\end{itemize}

\section{Simulation Results}
\label{sec:res}

We present results from Simulations 1 and 2 as follows. Transfer rates to relevant planets, the Sun, and ejection are listed in Table \ref{tab:transfer}. We also list transfer rates to the Moon for appropriate simulations. We discuss general trends observed in Simulation 3, and revisit the collision results in \S\ref{subsec:entry}.

\subsection{Simulation 1: Calibration Transfer Rates}

We explored different distributions of initial velocities, and found uniformly random sampling between $1-2\times v_{\rm esc}$ at the Hill sphere provided generally good agreement with \citet{Worth2013}. Our rates are within $1\sigma$ Poisson uncertainties of their rates in nearly all cases (note, \citet{Worth2013} do not provide rates from Venus). Differences may be attributed to ejection speed, launch position and velocity, integrator and close encounter implementations, and initial planet positions. 
The transfer rate from Venus to Earth ($10\%$) is encouragingly high. \citet{Melosh1993} present a larger transfer rate of $\sim 30\%$ from Venus to Earth, which we attribute to different starting conditions and integration method. About $40\%$ of particles in our simulations remain in orbit after 10 Myr, indicating that our transfer rates are lower bounds. We reran Simulation 1 for one million years, with inclusion of the Moon as an active particle. Most particles ($\gtrsim 80\%$) are still in orbit at the end of the simulation. Transfer rates from Venus to the Moon, Earth and re-accretion by Venus were  $0.03\%$, $4.48\%$ and $34.71\%$ respectively.

\setlength{\tabcolsep}{3pt}
\begin{table*}
  \centering
  \begin{tabular}{l r | c c c c c c c}
   Simulation & Origin & To Moon & To Sun & To Mercury & To Venus & To Earth & To Mars & Ejected \\
   \hline
   1 & 
     Venus & - & 2.05 $\pm$ 0.14 & 0.32 $\pm$ 0.06 & 48.55 $\pm$ 0.70 & 10.32 $\pm$ 0.32 & 0.18 $\pm$ 0.04 & 3.19 $\pm$ 0.18 \\
   1 & Earth & - & 2.43 $\pm$ 0.16 & 0.39 $\pm$ 0.06 & 13.49 $\pm$ 0.37 & 39.30 $\pm$ 0.63 & 0.21 $\pm$ 0.05 & 3.21 $\pm$ 0.18 \\
   1 & Mars  & - & 1.67 $\pm$ 0.13 & 0.03 $\pm$ 0.02 & 1.41  $\pm$ 0.12 & 2.62 $\pm$ 0.16 & 17.15 $\pm$ 0.41 & 2.45 $\pm$ 0.16 \\
   \hline
   W13 & 
     Earth & - & 1.5 $\pm$ 0.06 & 0.37 $\pm$ 0.03 & 13 $\pm$ 0.2 & 40 $\pm$ 0.3  & 0.18 $\pm$ 0.02   & 5 $\pm$ 0.1 \\
   W13 & Mars  & - & 1.3 $\pm$ 0.06 & 0.03 $\pm$ 0.009 & 1.5 $\pm$ 0.06 & 2.6 $\pm$ 0.08 & 16 $\pm$ 0.2 & 3.8 $\pm$ 0.1 \\
   \hline
   2 (a) ($V_i = 25$ \kms) &
     Venus & 0.10 $\pm$ 0.03 & 0.01 $\pm$ 0.01 & 0.03 $\pm$ 0.02 & 14.68 $\pm$ 0.38 & 4.03 $\pm$ 0.20 & 0.03 $\pm$ 0.02 & 0.03 $\pm$ 0.02 \\
   2 (a) ($V_i = 25$ \kms) & Earth & 0.08 $\pm$ 0.03 & $< 0.01$ & 0.03 $\pm$ 0.02 & 4.66 $\pm$ 0.22 & 17.24 $\pm$ 0.42 & 0.05 $\pm$ 0.02 & 0.03 $\pm$ 0.02 \\
   \hline
   \rowcolor{tablegray}[5pt][5pt] 2 (b) ($V_i = 30$ \kms) &
     Venus & 0.07 $\pm$ 0.03 & 0.11 $\pm$ 0.03 & 0.10 $\pm$ 0.03 & 10.94 $\pm$ 0.33 & 2.88 $\pm$ 0.17 & $< 0.01$ & 0.18 $\pm$ 0.04 \\
   \rowcolor{tablegray}[5pt][5pt] 2 (b) ($V_i = 30$ \kms) & Earth & 0.13 $\pm$ 0.04 & 0.51 $\pm$ 0.07 & 0.03 $\pm$ 0.02 & 2.98 $\pm$ 0.17 & 9.00 $\pm$ 0.30 & $< 0.01$ & 2.13 $\pm$ 0.15 \\
   \hline
   2 (c) ($V_i = 60$ \kms) &
     Venus & 0.01 $\pm$ 0.01 & 1.14 $\pm$ 0.11 & 0.16 $\pm$ 0.04 & 7.09 $\pm$ 0.27 & 2.08 $\pm$ 0.14 & 0.01 $\pm$ 0.01 & 8.44 $\pm$ 0.29 \\
   2 (c) ($V_i = 60$ \kms) & Earth & 0.03 $\pm$ 0.02 & 1.76 $\pm$ 0.13 & 0.05 $\pm$ 0.02 & 1.61 $\pm$ 0.13 & 3.23 $\pm$ 0.18 & 0.03 $\pm$ 0.02 & 15.62 $\pm$ 0.40 \\
  \hline
  \end{tabular}
  \caption{Transfer rates from our $N$-body simulations. W13 indicates results from \citet{Worth2013}. Rates are given in percentages of the total 10,000 particle population with Poisson uncertainties. Upper bounds indicate zero particle collisions in the simulation. Simulation 1 and those in \citet{Worth2013} were integrated for 10 Myr. Simulation 2 was integrated for 1 Myr.} \label{tab:transfer}
\end{table*}

\subsection{Simulation 2: Transfer of Impact Ejecta}

For the three considered impact scenarios, {corresponding to impact velocities of 25 \kms, 30 \kms\ and 60 \kms}, we resolve 10, 7 and 1 collisions of Venusian ejecta with the Moon, {respectively}. Transfer rates from Venus to Earth are also consistently high (of order several percent) across simulations. The orbital speed of Venus is approximately $35 \: \kms$, and ejecta in Simulation 2 have $v_{\infty}$ of order $7$ \kms. Ejecta with trajectories in the same direction as Venus' orbit attain elliptical orbits with semimajor axis $a \approx 1.3$ AU (this process is discussed further in \S\ref{sec:disc}). Ejection radially outward from Venus' orbit, as well as close encounters with Venus, may also place ejecta on orbits coinciding with Earth's. Particles with $v_{\infty}$ $\gtrsim 15 \: \kms$ along the orbit of Venus (for a total of $\gtrsim 50 \: \kms$) have sufficient speed to escape the Solar System. Since most particles remain in orbit, transfer rates to the Moon are lower bounds. 

As noted by \citet{Armstrong2002}, one may relate the number of impacts experienced by the Earth and Moon as the ratio of their gravitational focusing cross-sections:
\begin{equation}
    \frac{N_{\oplus}}{N_{\rm Moon}} = \frac{R_{\oplus}^2}{R_{\rm Moon}^2}\frac{v_{\infty}^2 + v_{\rm esc, \oplus}^2}{v_{\infty}^2 + v_{\rm esc, Moon}^2 + v_{\rm pot}^2}\, ,
\end{equation}
where $v_{\rm pot}$ is inversely proportional to the Earth-Moon separation and represents the excess in Moon collisions due to its proximity to Earth. We neglect it for the following discussion. We can expect the number of Moon collisions to be proportional to the number of Earth collisions, as the ratio above ranges from $18-255$ for residual velocities of $20$ \kms\ to $1$ \kms. In Simulations 2 (a) and 2 (b), the ratio of collisions with Earth and the Moon originating from Venus is about 40. This suggests particles interact with the Earth-Moon system at $\sim 7$ \kms\ on average. Fewer Earth and Moon collisions arise in Simulation 2 (c) (there was only one transfer from Venus to the Moon). Overall, a typical asteroid impact yields a 0.01-0.1 $\%$ Venus-to-Moon transfer rate over 1 Myr.

\begin{figure*}
    \centering
    \includegraphics[width=\textwidth]{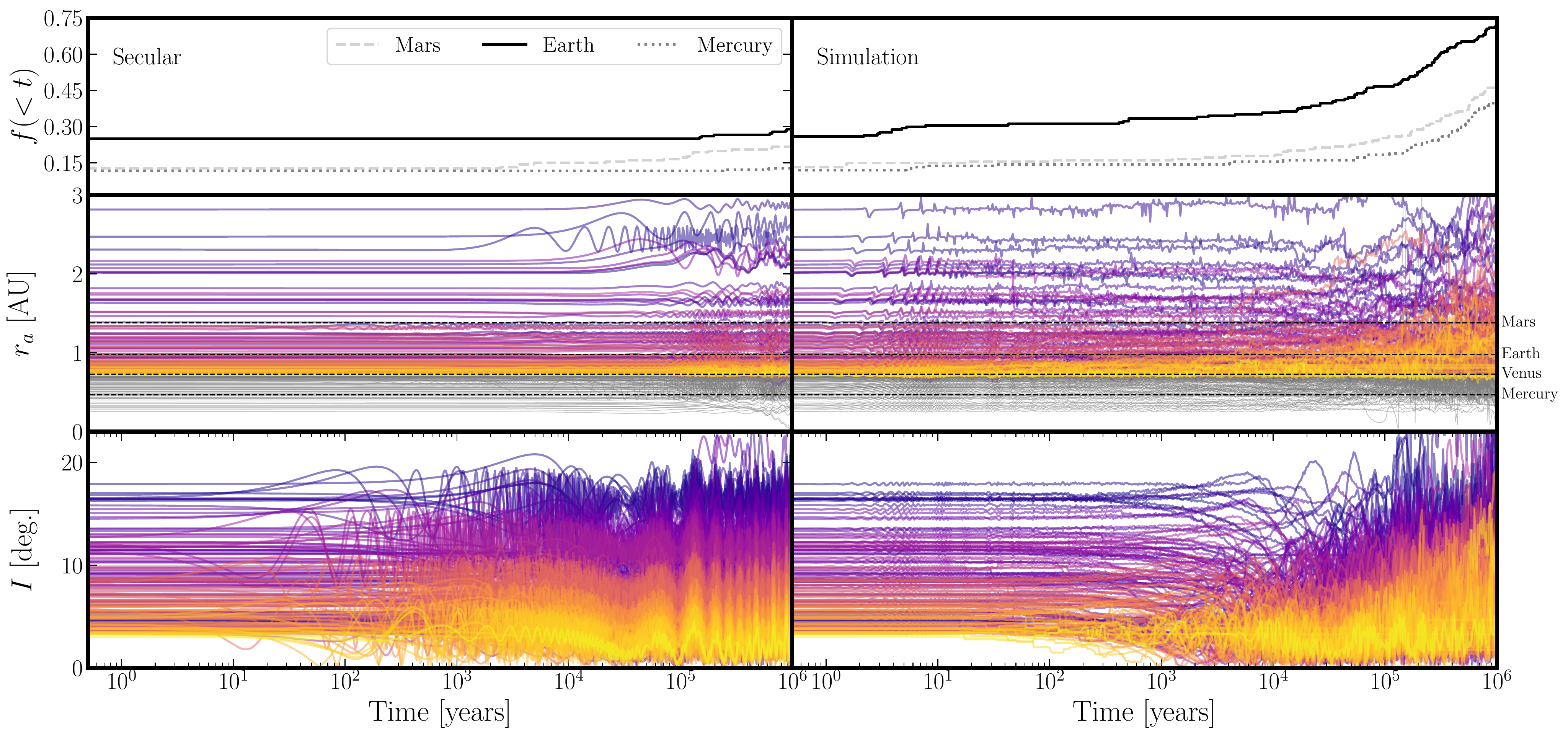}
    \caption{Orbital evolution of 180 ejecta particles from Simulation 2 (b), originating from Venus. Left and right panels correspond to theoretical secular evolution and numerical evolution respectively. {\it Top Panels}: cumulative fraction of particles which attain 1) pericenter less than Mercury's apocenter; 2) apocenter greater than Earth's percienter; and 3) apocenter greater than Mars' pericenter.  {\it Center Panels}: apocenter for each ejecta particle. Color denotes the residual velocity at ejection (brighter colors correspond to lower velocities). Light gray tracks indicate the evolution of pericenter. Dashed horizontal lines indicate the threshold necessary to cross the orbits of Mercury, Venus, Earth, and Mars. A significant number of particles attain Earth-crossing orbits past $\sim10^5$ years. {\it Bottom Panels}: evolution of orbital inclination, which shows long-term similarities between the secular and numerical evolutions.}
    \label{fig:evo}
\end{figure*}

Our simulations lead to a remarkable conclusion: impacts onto Venus and Earth transport material to the Moon with comparable efficiency {for each considered impact speed}. Although this conclusion is new, it can be deduced from previous estimates. Our secular analysis (\S\ref{sec:secular}) predicts a large fraction of ejecta launched from the leading hemisphere of Venus will attain Earth-crossing orbits. Indeed, in Simulation 1 we find that while re-accretion by Venus is the most common trajectory, Earth acts as a particularly effective attractor and accretes up to $10\%$ of Venusian material. Also, the reciprocal Earth-to-Venus transfer rate is $13\%$, matching findings of \citet{Worth2013}. The proximity and similarity of the two planets suggests the two rates should be comparable. In Simulation 2 (b), most ejecta attain elliptical, Earth-crossing orbits (Figure~\ref{fig:evo}). Most have insufficient energy to reach Mars and do not cross Mercury's orbit either; these two planets also have considerably less mass than Earth and Venus. In the case of no focusing (particles with infinite interacting velocity), Earth should experience $13.45\times$ as many collisions as the Moon. In the opposite limit, as the interacting velocity tends to zero, Earth should experience $297.8\times$ as many collisions. Under this argument, we should expect a Venus-to-Moon transfer rate between $0.03-0.7\%$. All 1 Myr transfer rates lie in this range except for the 60 \kms\ impactor scenario. {We do note the simulated Moon transfer rates have large uncertainties, which might be improved with longer simulations and more ejecta particles.}

What mechanism is responsible for the transport of ejecta to Earth's orbit? Figure~\ref{fig:evo} compares the orbital evolution of a subset of particles to the evolution predicted by secular theory. The diversity of tracks under secular evolution is explained as follows: particle orbital elements oscillate rapidly if they have semi-major axes near that of a planet (i.e. near a precession rate singularity). Large oscillations occur when semi-major axes lie near eigenfrequencies of either eccentricity or inclination. The numerical evolution has long-term qualitative similarities to the secular evolution, but it does not display either of the foregoing phenomena. Rather, there is small-scale high-frequency structure (mostly at the period of Venus' orbit). This trend demonstrates that resonances and close-encounters with Venus prohibit growth of secular resonances, and eventually lead to chaotic scattering of particles with a characteristic time scale exceeding $\sim100,000$ years. The Earth-Moon system is sufficiently nearby to receive the influx of ejecta whose orbits are excited by Venus in this manner.

\subsection{Simulation 3: Enlarged Earth Particle}

\begin{figure*}
    \centering
    \includegraphics[width=\textwidth]{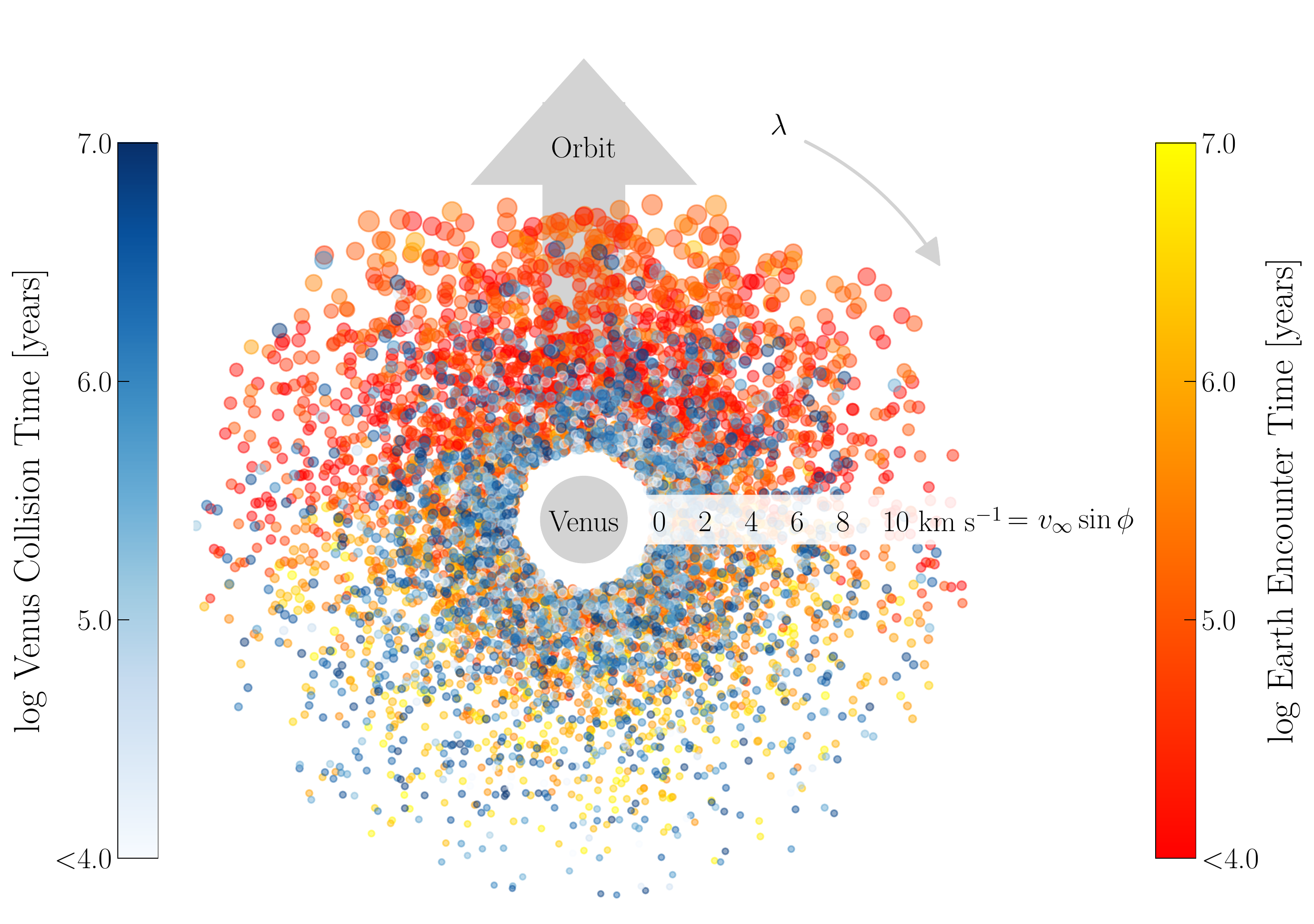}
    \caption{Initial conditions of ejecta launched in Simulation 3 (corresponding to a 35 \kms\ impactor), color-coded by the time they reach their destination. Each particle is launched from a randomly drawn longitude $\lambda$ and latitude $\phi$, with residual velocity $v_{\infty}$. The distance of particles from the center of the figure  is proportional $v_{\infty}\sin{\phi}$, the component of the residual velocity lying in the orbital-plane. Particle size is proportional to initial semi-major axis, which ranges from $0.45 < a_0 < 2.84$. Red-to-yellow hued particles collide with the enlarged Earth particle, which is classified as an encounter with the Earth-Moon system. White-to-blue hued particles are re-accreted by Venus. Particles launched from the leading face of Venus are immediately placed on Earth-crossing orbits.}
    \label{fig:launch}
\end{figure*}

Results from Simulation 3 provide insight into the evolution of ejecta orbits, and how they come to intersect Earth's. We record a significant number of collisions between Venusian ejecta and the enlarged Earth particle ($42\%$). The second most common outcome was re-accretion by Venus ($22\%$). The initial conditions and destinations of ejecta are shown in Figure~\ref{fig:launch}. There is no strongly preferred launch direction or energy for collision with the enlarged Earth particle. However, particles launched from the leading face of Venus immediately acquire Earth-crossing orbits, and are more likely to encounter the Earth on short timescales ($<10,000$ years). Particles launched from the trailing face of Venus do not have sufficient energy to reach Earth; however, as discussed in the previous subsection, repeated close encounters or resonances with Venus are responsible for increasing their semi-major axes. There is no apparent correlation between launch velocity and collisions with Venus, or the times of collision. Over $80\%$ of all collisions with Venus take place more than 100,000 years after launch.

\section{Ejecta Flux Analysis}
\label{sec:ej}

\subsection{Case Study: Nominal Asteroid Projectile}

\begin{figure}
    \centering
    \includegraphics[width=0.45\textwidth]{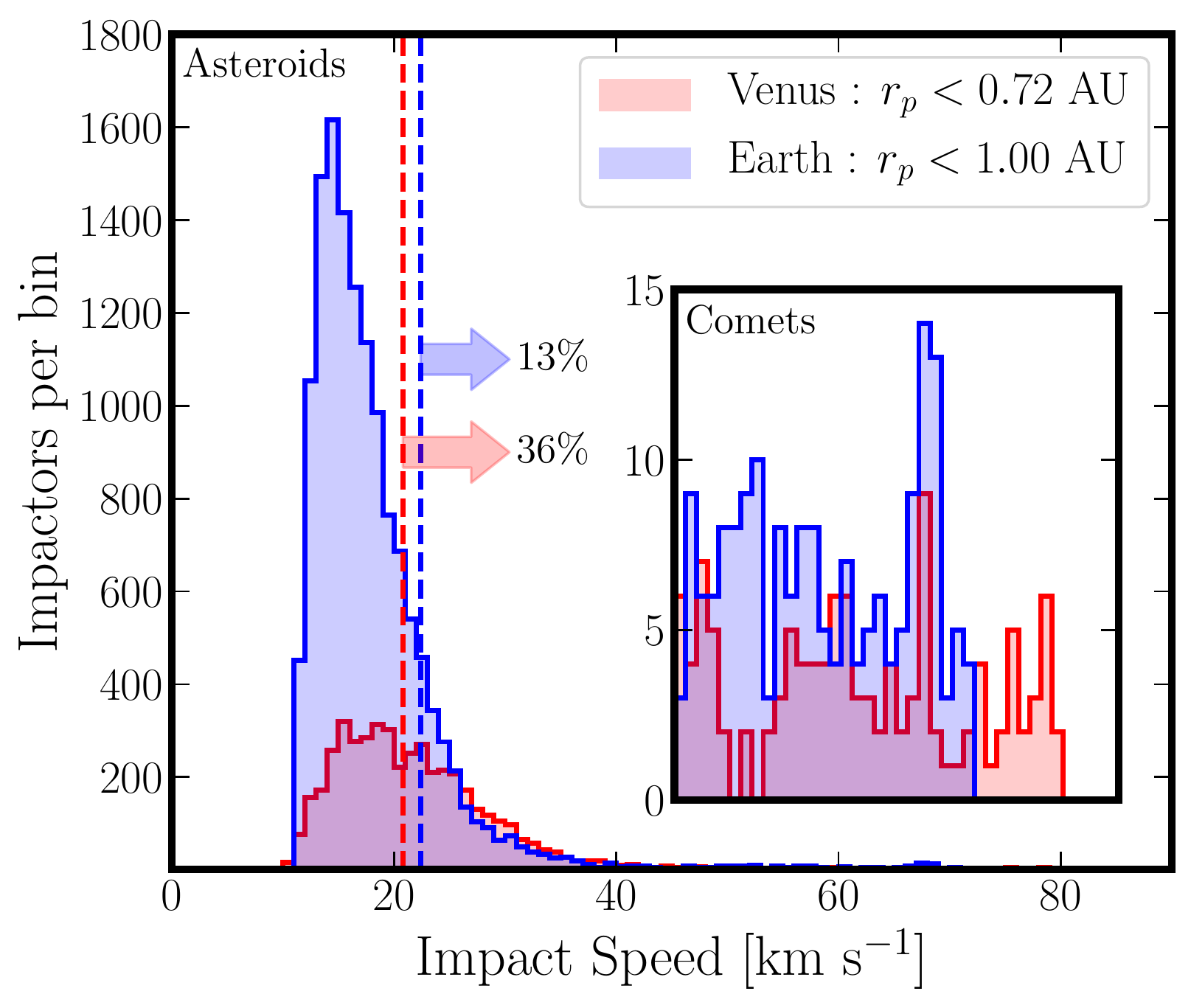}
    \caption{Distribution of potential collision speeds with Earth and Venus, based on asteroids and comets in the JPL Small Bodies Database. Only projectiles with a perihelion distance within the planet's orbit are considered. This amounts to 15330 and 6299 objects for Earth and Venus respectively. Histograms have 1 \kms\ bins. The inset shows a zoom-in of the high-velocity regime (45 to 85 \kms), mostly comprising comets, which there are far fewer number of than asteroids. Vertical dashed lines mark twice the escape velocity of Venus (red) and Earth (blue). A significantly higher fraction of possible impacts with Venus occur at speeds above twice the escape velocity (36$\%$) as compared to Earth (13$\%$), owing to Venus' proximity to the Sun.}
    \label{fig:impdistr}
\end{figure}

Careful consideration must be given to projectile characteristics given the limits on spalled material. For example, the Chicxulub asteroid had diameter $D \sim 10$ km and impact velocity $V_i \sim 20$ \kms\ \citep{Alvarez1980, Schulte2010}, and despite all the vicissitudes it created for surface fauna, its impact velocity was insufficient to launch spall fragments at the escape velocity of Venus or Earth since $ v_{\rm esc} > V_i/2 \geq v_{\rm ej}$. A projectile needs $V_i \gtrsim 22$ \kms, or more considering atmospheric drag on ejecta. Heliocentric orbits of asteroids between Mars and Jupiter yield impact velocities around $15-20$ \kms\ for Earth (note, impact velocity is lower bounded by the planet's escape velocity). On average, asteroids hit Venus with $24\%$ higher speed, since Venus lies deeper in the Sun's gravitational well. This trend is shown in Figure~\ref{fig:impdistr} in which we plot our calculated distribution of impact speeds for asteroids and comets in the JPL Small Bodies Database\footnote{https://ssd.jpl.nasa.gov}. Impact speeds are approximate, and based on vector addition of velocities between the planet and projectile at the planet's orbital radius. The planet's escape velocity was then added in quadrature. Projectiles in the two distributions were limited to those with perihelion within the planet's orbit. Only $13\%$ of potential Earth collisions exceed twice the escape velocity, whereas $36\%$ of potential Venus collisions do. Under more conservative assumptions that spall material achieves only $35\%$ of the impactor speed, and considering atmospheric drag, \citet{Melosh1988} demonstrate that at an impact speed of at least $30$ \kms\ is necessary to eject material from Earth. About $3\%$ and $9\%$ of possible collisions with Earth and Venus respectively exceed $30$ \kms. Our findings are in excellent agreement with \citet{Steel1998}, who find only a few percent of Earth impacts exceed $30$ \kms\ and $25\%$ of impacts exceed $20$ \kms. Figure~\ref{fig:impdistr} also shows that significantly fewer small bodies cross Venus' orbit compared to Earth's. \citet{Shoemaker1991} estimate Venus experiences $0.86\times$ the collisions of Earth (given $59\%$ of Earth crossing asteroids also cross Venus' orbit, and the mean collision rate is $1.45\times$ higher for Venus). Since the impact velocities are higher for Venus, the cratering rate roughly matches that of Earth \citep{Shoemaker1991, leFeuvre2008}. 

{The ancient impact velocity distribution is difficult to ascertain and is usually estimated with models of the early Solar System. Collisions between asteroids today occur at $\sim 5.3$ \kms, whereas models indicate collisions prior to the formation of Jupiter occurred at $\sim 6$ \kms, partially excited by planetary embryos \citep{Bottke2005}. It is possible that velocities were higher during LHB. Currently, Lunar asteroids impacts occur at $\sim 14$ \kms. Models by \citet{Bottke2012} indicate $\sim 9$ \kms\ Lunar impact speeds prior to LHB, and $\sim 20$ \kms\ during LHB. If this trend translates into higher impact speeds with terrestrial planets, then more spall material is ejected. However, given the uncertainty surrounding actual conditions $>3.5$ Gyr ago, and for the sake of simplicity, we assume the current impact velocity distribution holds throughout the Solar System's history. }

Our study mainly concerns asteroid impacts; however we take a brief aside to review the potential influence of comets. Indeed, \citet{Steel1998} suggest that high impact speeds may lend special importance to spallation from comet impacts. The present-day flux of comets to Venus should be comparable to that experienced by Earth \citep{Shoemaker1987}; even during LHB, \citet{Rickman2017} estimate Earth may have encountered only $\sim25\%$ more cometary projectiles than Venus. \citet{Weissman2007} state a long-period comet impact rate of $2.6\times10^{-8}$ yr$^{-1}$, or one per 38 Myr. For Halley-type comets (periods between 20 and 200 years), the rate is as high as $1.9\times10^{-8}$ yr$^{-1}$, or one per 52 Myr. These two classes have most-probable Earth-impact velocities of 53.5 \kms\ and 51.3 \kms, respectively (Jupiter-family comets have insufficient speed to launch spall material). A powerlaw or broken-powerlaw model is often used to model the size distribution of comets, which ranges from less than 1 km to up to 30 km \citep{Weissman1996, Meech2004, Fernandez2012}. \citet{Artemieva2004} show that comets are relatively inefficient at ejecting spall material, due to their reduced equivalent depth which scales as $(\rho_p\rho_t)^{0.5}$. Comets have typical bulk density of 0.6 ${\rm g \: cm^{-3}}$ \citep{Weissman2004}, versus 2.86 g cm$^{-3}$ basalt density. The low density of comets propagates into the coupling constant $C_V$, and reduces particle speeds by a factor of $\sim4$. Indeed, \citet{Artemieva2004} find comet impacts on Mars eject less material than much slower asteroid impacts. 

One additional consideration is interstellar objects such as ‘Oumuamua \citep{Meech2017} or Comet 2I/Borisov \citep{Guzik2020}. Their detections provide some constraints on number densities of similar objects (e.g. ‘Oumuamua-like objects have ${n}_{\rm IS} \approx 0.2 {\,\rm AU}^{-3}$ per \citet{Do2018}), making them plausible, high-velocity impact projectiles. Accounting for gravitational focusing, the \citep{Do2018} space density implies an impact rate, $\Gamma=n\sigma v$, of roughly once per $10^{8}\,{\rm yr}$ for 'Oumuamua-like objects. 

It is challenging to empirically verify the spallation mechanism for physically large projectiles, especially since pressures usually exceed those intended by the hydrodynamical model \citep{Melosh1984}. However, it has largely held true in small-scale experiments \citep[e.g.][]{Gratz1993}. Given that Venus encounters many asteroids with sufficient impact velocity to launch spall material, we do not consider comets or interstellar objects any further. As done in previous literature \citep{Melosh1984, Mileikowsky2000} we restrict analysis to vertical impactors. However oblique impacts with Earth and Venus warrant future investigation with hydrodynamical modeling, in particular because oblique impacts are more common, and can launch up to $100\times$ more spall material for Martian impacts \citep{Artemieva2004}. 

Earth currently experiences asteroid collisions at a rate $\sim4.3\times10^{-6}\:{\rm yr}^{-1}$ \citet{Shoemaker1990} {for asteroids with diameter $\gtrsim$ 1 km. This rate agrees within a factor of $\sim 2$ of more current estimates \citep{Bland2006}}. For exceptionally large asteroids (diameter $\gtrsim$ 8 km), the collision rate is $\sim3\times10^{-9}\:{\rm yr}^{-1}$. The differential size distribution of asteroids between $1-100$ km contains subtle features \citep{Bottke2005}; however, a power law of index $-3.5$ is a reasonable approximation \citep{Dohnanyi1969} (the region $3-30$ km is better described by an index $-2.9$ per \citet{Brien2003}). To reduce the number of free parameters, we nominally assume an asteroid projectile with diameter $8$ km, impacting vertically at $30$ \kms.
    
{Ancient} Venusian ejecta may comprise igneous rocks such as basalt, andesite and granite. {Indeed, there are some indications of present-day silicic rocks \citep{Hashimoto2008, Basilevsky2012}, which may have been in even greater abundance prior to Venus' resurfacing.} Tensile strengths are typically between $0.01-0.1$ GPa \citep{Melosh1992, Mileikowsky2000}, with some dependence on ambient conditions \citep{Schultz1993}. We take the sound-speed in rock to be $c_L \approx 6$ \kms\ \citep{Grady1979}, and basalt density $\rho_t \approx 2.86$ g cm$^{-3}$ \citep{Melosh1989}. For a projectile of equal density, we find the total mass of material ejected at $v_{\rm ej} \geq 10.36 \: \kms$ and shocked below 40 GPa is $m_{\rm ej}/{m_i} = 0.049$ (Table~\ref{tab:spall}). Unlike the analysis presented in \citep{Mileikowsky2000}, we include rock that is shocked above 1 GPa. The ejected mass is $m_{\rm ej} = 3.7 \times 10^{13}$ kg. Taking a tensile strength of 0.1 GPa \citep{Mileikowsky2000}, the typical Grady-Kipp fragment size is about 0.6 m. We calculate $N_{\rm ej} \simeq 1\times10^{11}$ particles are ejected. We caution that this is an order-of-magnitude estimate. However, it follows reasonably from average properties of the projectile and target, and hydrodynamical impact theory. Atmospheric drag increases the required launch velocity \citep[e.g.][]{Schultz1979}, effectively preventing ejection of any material from modern-day Venus. Drag is less dramatic for Earth-like atmospheres, and \citet{Melosh1988} note large projectiles clear out the atmosphere directly above the impact, reducing drag. 

\setlength{\tabcolsep}{3pt}
\begin{table}
  \centering
  \begin{tabular}{c | c c c c c c}
   $V_{i}$ (Venus) & $<1$ GPa & $<10$ GPa & $<40$ GPa & $<70$ GPa \\
   \hline
   25 \kms & 6.36E-04 & 6.36E-03 & 2.54E-02 & 4.45E-02 \\
   30 \kms & 1.23E-03 & 1.23E-02 & 4.92E-02 & 8.61E-02 \\
   60 \kms & 3.52E-03 & 3.52E-02 & 1.41E-01 & 2.46E-01 \\
   \hline
   \hline
   $V_{i}$ (Earth) & $<1$ GPa & $<10$ GPa & $<40$ GPa & $<70$ GPa \\
   \hline
   25 \kms & 3.47E-04 & 3.47E-03 & 1.39E-02 & 2.43E-02 \\
   30 \kms & 9.04E-04 & 9.04E-03 & 3.62E-02 & 6.33E-02 \\
   60 \kms & 3.00E-03 & 3.00E-02 & 1.20E-01 & 2.10E-01 \\
  \end{tabular}
  \caption{Fraction of impactor mass ejected at escape velocity or greater ($m_{\rm ej}/{m_i}$), under various impact conditions. The top section corresponds to impacts with Venus, the bottom to Earth. For each impact velocity (rows), the fraction given is that of mass shocked below pressure $P_{\rm max}$ (columns).} \label{tab:spall}
\end{table}

We estimate the number of ejecta fragments (originating from a planet denoted X) which impact the Moon as meteorites due to a single asteroid impact by:
\begin{equation}
    N_{\rm moon} = N_{\rm ej} \times f_{\rm X-Moon}\,.
\end{equation}
 Given $N_{\rm ej} \simeq 1\times10^{11}$, and a lower bound $f_{\rm V-Moon} = 7\times10^{-4}$, we calculate $N_{\rm moon} = 7\times10^7$. The transfer efficiency $f_{\rm V-Moon}$ corresponds to ejecta from Simulation 2 (b), which had $v_{\infty}$ between $0 - 11$ \kms. Given $R_{\rm moon} = 1737$ km, our result corresponds to 2 Venusian ejecta per square kilometer on the Moon's surface, or about 680 kg km$^{-2}$ from a single asteroid impact. Our transfer rates are within the same order of magnitude for low and high velocity ejecta, modelled by Simulations 2 (a) and 2 (b). The true quantity of ejected material depends on the mass and speed of the impactor, and the total amount of material on the Moon depends on the frequency of impactors. We take these factors into consideration in the following subsection. 

In Simulation 3, we excluded the Moon as a massive particle and set the radius of the Earth to the Moon's orbital separation. We found very high transfer rates, of up to $\sim 42\%$ from Venus to the enlarged Earth particle for 10 Myr integrations. A proxy for the impact rate to the Moon is:
\begin{equation}
    N_{\rm moon} = N_{\rm ej} \times f_{\rm X-E} \times \frac{A_{\rm moon}}{A_{\rm col}} \,,
\end{equation}
where $A_{\rm moon}/A_{\rm col} = 2.4\times10^{-5}$ is the ratio of the Moon's cross-sectional area to the surface area of a collision sphere centered on the Earth enclosing the Moon's orbit, and $f_{\rm X-E}$ is the transfer rate to the enlarged particle. {That is, $A_{\rm moon} = 4\pi R_{\rm moon}^2$, where $R_{\rm moon}$ is the Moon's radius, and $A_{\rm col} = 4\pi \eta^2$, where $\eta$ is the orbital separation of the Moon}. The small fractional area occupied by the Moon yields a significantly lower Venus-to-Moon transfer rate than found in Simulation 2. This result indicates that repeated close-encounters are important. That is, particles which come within the Moon's separation from Earth but miss a collision have a non-negligible chance of future close-encounters.

\subsection{Estimating Venusian Material on the Moon}

We use a Monte-Carlo analysis to estimate the abundance of Venusian material on the Moon, taking into account potential impact and water loss histories. We assume that Earth and Venus experience identical fluxes of asteroids. {The impact rate is obtained} by integrating the Lunar Sawtooth Bombardment rate of \citet{Morbidelli2012}, and multiplying it by the ratio of Earth's and the Moon's geometric cross-sections. {An advantage of the \citet{Morbidelli2012} model is its calibration to siderophile analyses, which suggest a total impact flux of $\sim 3.5 \times 10^{19}$ kg. For contrast, we also tested the smooth post-accretionary decline model of \citet{Neukum1994}, which yields $\sim 4\times$ more spalled ejecta, mainly because it predicts a higher impact rate prior to 4.1 Gyr ago}. At times longer than $~1$ Gyr after the Solar System's formation, the impact rate is well described by 430 per 100 Myr \citep{Shoemaker1990}. 
We sample the duration for which Venus retained water (which we treat as synonymous to the duration over which it harbored a low-mass atmosphere):
\begin{equation} 
    t_{\rm ret}{\rm (Gyr)} \sim \mathcal{U}(0.01, 1.50)\, .
\end{equation}
This assumes Venus had water at formation $4.5$ Gya, and places a uniform prior on the water loss time between 10 Myr and 1.5 Gyr after formation. While water loss may have occurred as late as 0.7 Gya \citep{Way2016}, the vast majority of impacts occurred in the first $\sim1$ Gyr of the Solar System. After $t_{\rm ret}$, we assume no more material may be be ejected due to drag from the thick atmosphere. {Prior to $t_{\rm ret}$, the model assumes Venus' atmosphere was sufficiently thin such that the effects of drag may be ignored. In reality, drag may increase the minimum impact speed necessary to eject lightly shocked material; however this effect is limited since the asteroid evacuates the atmospheric column in the vicinity of the impact site \citep{Melosh1988, Cataldi2017}. Our model does not include various stages of the atmosphere's evolution, which may impart various degrees of drag.} {We sample the number of arrivals from a Poisson distribution. Its mean is set to the integral of the impact history up to $t_{\rm ret}$. If $t_{\rm ret} > 1$, then we separate the calculation into epochs before and after 3.5 Gyr ago for reasons discussed below.} For each collision, we sample an impact velocity from the distribution for present-day asteroids, shown in Figure~\ref{fig:impdistr}. Next, we calculate the projectile mass using a bulk density of 2.86 ${\rm g \: cm^{-3}}$ and diameter drawn from a power-law cumulative distribution function:
\begin{equation}
    N(>D) \propto D^{b} \,.
\end{equation}
We choose an index uniformly at random between $-1.0 > b > -2.5$ (corresponding to a differential power-law index of between $-2.0$ and $-3.5$), and we restrict sampling of the distribution from $1-8$ km. We then use Equation~\ref{eqn:ratio} to estimate the mass of ejecta shocked below 40 GPa. Finally, we apply a transfer rate of material to the Moon drawn from a uniform distribution between 0.01$\%$ and 0.1$\%$ (span of simulated 1 Myr transfer rates). If the sampled impact velocity is below twice the escape velocity, then the impact does not transfer any material. 

{The Monte-Carlo analysis yields a distribution of possible surface densities of Venusian material on the Moon. The 16$^{\rm th}$, 50$^{\rm th}$ and 84$^{\rm th}$ percentiles of the distribution are $6.2\times10^5$, $9.5\times10^5$, and $1.7\times10^6$ kg km$^{-2}$, respectively. The majority of this material originates from impacts prior to 3.5 Gyr ago. As such, it will not be uniformly mixed into the Lunar regolith. It is well-understood that various depths trace different epochs of the Moon's history \citep[e.g][]{Hiesinger2006} as discussed in the following section. We adopt a simple stratification model to estimate the abundance of Venusian material. It is assumed that impacts prior to 3.5 Gyr ago are mixed within the megaregolith to depths of 1000 meters, whereas subsequent impacts are mixed within the surface regolith down to 10 meters (in reality the transition is less distinct, and the depth of upper regolith varies across the Moon's surface). The median abundance of lightly-shocked Venusian material in the deep Lunar megaregolith is 0.3 ppm. One-third of samples have $t_{\rm ret} > 1$. Of these Monte-Carlo samples, the median abundance of Venusian material is 0.2 ppm. The two abundances are similar owing to the fact that, while the post-LHB impact rate is $\sim100$ times lower, it is mixed to depths about the same factor shallower.}

\subsection{Estimating Earth Material on the Moon}

Earth ejecta will constitute a significant fraction of the non-Lunar material on the Moon's surface. For the purposes of distinguishing Venusian ejecta from Earth ejecta, we thus also investigate the flux of Earth material to the Moon. This quantity is discussed at length by \citet{Armstrong2002}. We briefly review their analysis, and then provide an independent estimate from a Monte-Carlo analysis similar to that in the previous subsection. \citet{Armstrong2002} provide transfer efficiencies for three avenues of transport from Earth to the Moon: direct, for particles traveling below Earth's escape velocity; orbital, for particles which enter a heliocentric orbit and re-enter the Earth-Moon system; and lucky shots, for fast particles that collide with the Moon on their way out. Transfer rates are consistently higher for smaller Moon orbital separation, which, at a time 4 Gyr ago is estimated to have been about one-third of its current value \citep{Zharkov2000}. While they find an overall abundance of Earth material on the Moon of 7 ppm, they note that most of it is not spalled and is therefore probably shock-altered. More recently, \citet{Armstrong2010} find a lower abundance of 1-2 ppm heavily-shocked Earth material, with higher concentrations on the leading side of the Moon. 

We repeated our Monte-Carlo analysis described above to independently estimate the abundance of lightly-shocked Earth material on the Moon. We allowed impacts over {4.5} Gyr, and sampled the Earth-Moon transfer rate from a uniform distribution between 0.03$\%$ and 0.13$\%$. {The median abundance in the deep megaregolith is 0.1 ppm, and reaches 0.7 ppm in the surface regolith}. We note a small portion of lightly shocked Earth material on the Moon arrives from direct transfer. We estimate this fraction as follows for Earth-Moon separations $(\eta)$ of 21.6, 41.0 and 60.3 \rearth, a subset of those considered by \citet{Armstrong2002}. The minimum ejection velocity ($v_{\rm min}$) required to reach these separations are 10.94, 11.06 and 11.11 \kms\ respectively, notwithstanding the Moon's gravity. For a nominal 30 \kms\ vertical impactor, the fractions of impactor mass shocked below 40 GPa and ejected at $v_{\rm min}$ or greater ($m_{\rm ej}/m_{\rm i}$, given by Equation~\ref{eqn:ratio}) are 0.0399 0.0382, 0.0374 for each value of $\eta$. The fraction of mass ejected at Earth's escape velocity of 11.2 \kms\ is 0.0362. These calculations highlight that only a small fraction of material ejected fast enough to reach the Moon is not ejected from the Earth-Moon system ($3-9\%$). Transfer rates are as high as $0.014\%$ \citep{Armstrong2010}. Therefore the mass of material which lands on the Moon is smaller than that delivered from orbit by our nominal 30 \kms\ impactor. Some asteroids may impact below twice the escape velocity but still fast enough for spall material to reach the Moon. At exactly twice the escape velocity, a 22.4 \kms\ impactor ejects 0.0031 of its mass at sufficient velocity to reach the Moon at its closest separation. However, only a small fraction of asteroids ($\sim 2\%$) impact at velocities in this narrow range. The amount of directly transferred, lightly-shocked Earth material is thus one to two orders of magnitude less than that transferred from orbit. 

The felsic clast found in the Apollo 14 sample \citep{2019Bellucci} {could provide a broad,} independent estimate of the abundance of lightly-shocked terrestrial material on the Moon. The clast weighs 1.8 g whereas Apollo 14 returned 43 kg of material \citep{Wilshire1972}, which would point toward a naive abundance of about 42 ppm. {This abundance is hard to reconcile when our Monte-Carlo analysis indicates it should be at least an order-of-magnitude lower. A Lunar origin seems more likely \citep{Warren2020}.}

\subsection{Tracing the Water History of Venus}

\begin{figure}
    \centering
    \includegraphics[width=0.45\textwidth]{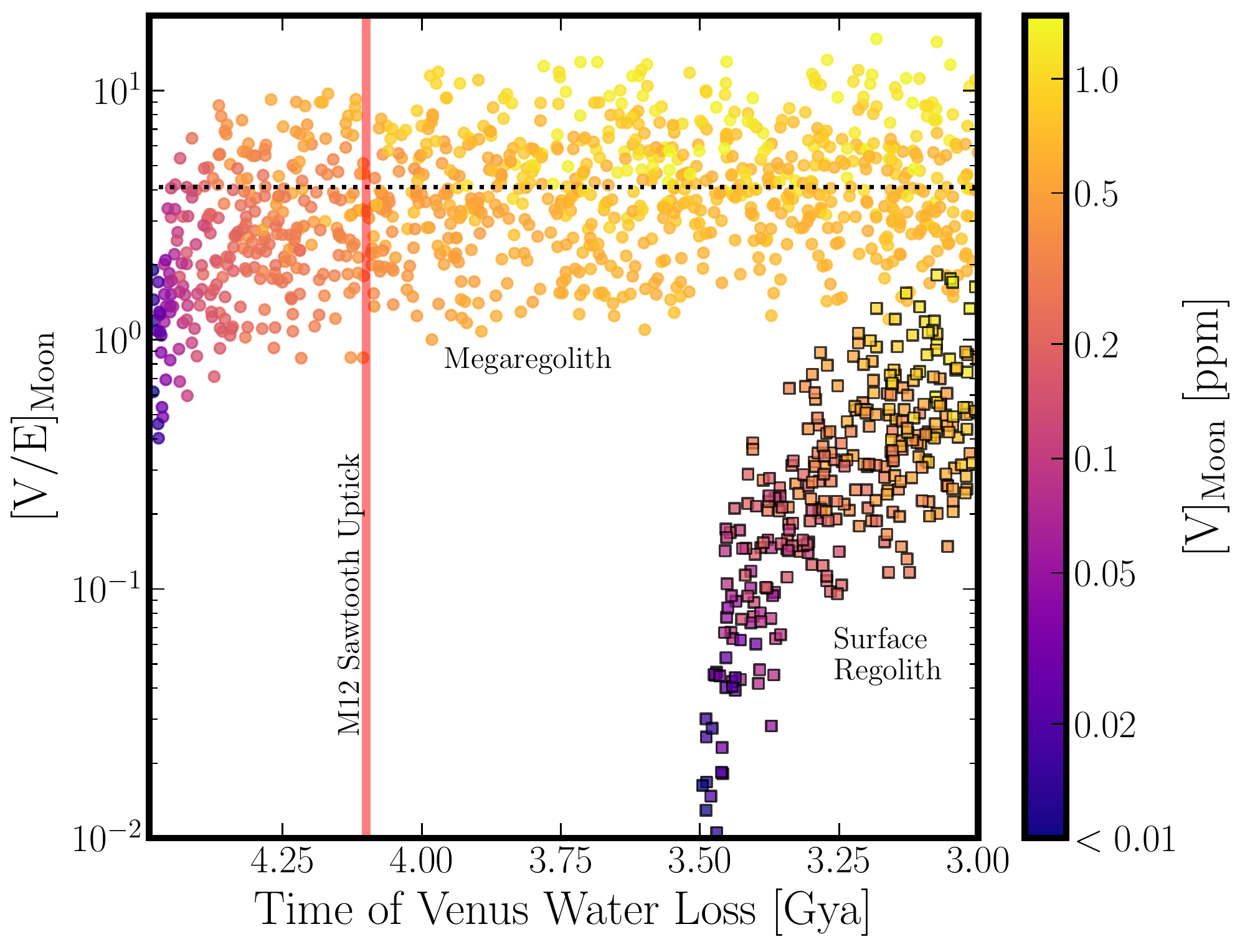}
    \caption{Monte-Carlo analysis of the abundance ratio of Venus and Earth material on the surface of the Moon, {in both the surface regolith and deeper megaregolith}. This distribution is marginalized over impact-related variables to reveal a function of when Venus lost the bulk of its water content. The ratios approach about 4.1 (black dotted line). The vertical red {line} is located at the {bombardment rate uptick} \citep{Morbidelli2012}. The colorbar indicates the abundance of lightly-shocked Venusian material, according to uniformly random drawn mixing depths.}
    \label{fig:time}
\end{figure}

By combining the analyses above, we obtain {the abundances of Earth and Venusian material on the Moon under various assumptions of number of arrivals, their impact speeds and radii, and Venus-to-Moon and Earth-to-Moon transfer rates. Abundances are based on mixing depths drawn uniformly at random between 500 and 1500 meters for the megaregolith calculation, and between 5 and 15 meters for the surface regolith. We now consider the dependence of abundance on water retention time, and plot this one-dimensional function} in Figure~\ref{fig:time}. 

Remarkably, even if Venus lost its water shortly after its formation, we still expect the ratio of lightly shocked Venusian material on the Moon to exceed that of similarly unaltered terrestrial material by a factor of more than two. This conclusion seems robust, as it is determined by fixed variables such as Earth's and Venus' escape velocities, their relative impact flux, and their respective impact velocities; it is independent of the absolute impact flux, and is relatively robust to low-frequency impactors (e.g. comets, interstellar objects). The main source of uncertainty is the transfer rate from Venus and Earth to the Moon for impact ejecta with various residual velocities, which we have constrained in \S\ref{sec:res}. Our analysis shows: 1) that if Venus had a {long-lived} thin atmosphere, then detectable quantities of Venusian rock are likely present in the current inventory of Lunar material (which consists of the samples returned by the Apollo Missions and the collection of Lunar meteorites); and 2) if we are able to obtain a collection of Earth and Venus samples on the Moon, the ratio of their abundances {may offer some loose constraints on Venus' evolution}. As a consequence of the high early rate of impacts, the {surface regolith} profile in Figure~\ref{fig:time} gives tight dating of water-loss times occurring {after the end of the Lunar cataclysm}. {However, these inferences require reliable abundance estimates of both Venusian and Earth material on the Moon, which would likely only be possible with hundreds or thousands of samples obtained at various locations and regolith depths. Also, it might be necessary to obtain samples from the megaregolith, which is currently infeasible.}

For long-lived oceans on Venus, the V/E abundance ratio saturates at $[{\rm V}/{\rm E}]\sim 4.1$ {in the megaregolith}, which can be approximately explained as follows. For asteroids that collide with speeds at least $2\times v_{\rm esc}$, the median collision speed is $\sim 25$ \kms, for collisions with Earth and Venus. Such a collision with Venus ejects $1.8\times$ more spall material than a collision with Earth, due to Venus' lower escape speed. Also, Venus experiences about  $\sim 3\times$ more spall-ejecting collisions than does Earth. The Venus-to-Moon transfer rate is slightly lower than the Earth-to-Moon transfer rate, such that these quantities yield a net ratio of ${\rm [V/E]}_{\rm Moon} \sim 4$ in the long-term. This ratio is dominated by collisions from the beginning of the Solar System, and is qualitatively explained by the locations of Venus and Earth. Venus experiences more spall-ejecting collisions than Earth, and spall layers are ejected with residual velocities that are serendipitously efficient for transfer to Earth. Venusian impactors lend energy to transport Venusian material out of the Sun's potential well. Earth however does not experience many collisions of sufficient speed. {Our model only considers Venusian material in the surface regolith if it arrived between 3.5 and 3.0 Gyr ago, whereas Earth ejecta has accumulated since 3.5 Gyr ago. Therefore the ratio is lower in the surface layer.}

The ratio of Venus to Earth material may be even higher than our Monte-Carlo estimate. For example, Venus may have lost its water content without experiencing a runaway greenhouse effect \citep{Abe2011}, thus maintaining a thin atmosphere until as recently as 1 Gya. If liquid water covered $\sim 70\%$ of Earth from an early age, then spall ejections would be further restricted to collisions with dry terrain or shallow water. Spall ejection from a dry yet temperate Venus, however, would not have this restriction. 

\section{Discussion}
\label{sec:disc}

\subsection{Lunar Entry and Regolith}
\label{subsec:entry}

\begin{figure*}
    \centering
    \includegraphics[width=\textwidth]{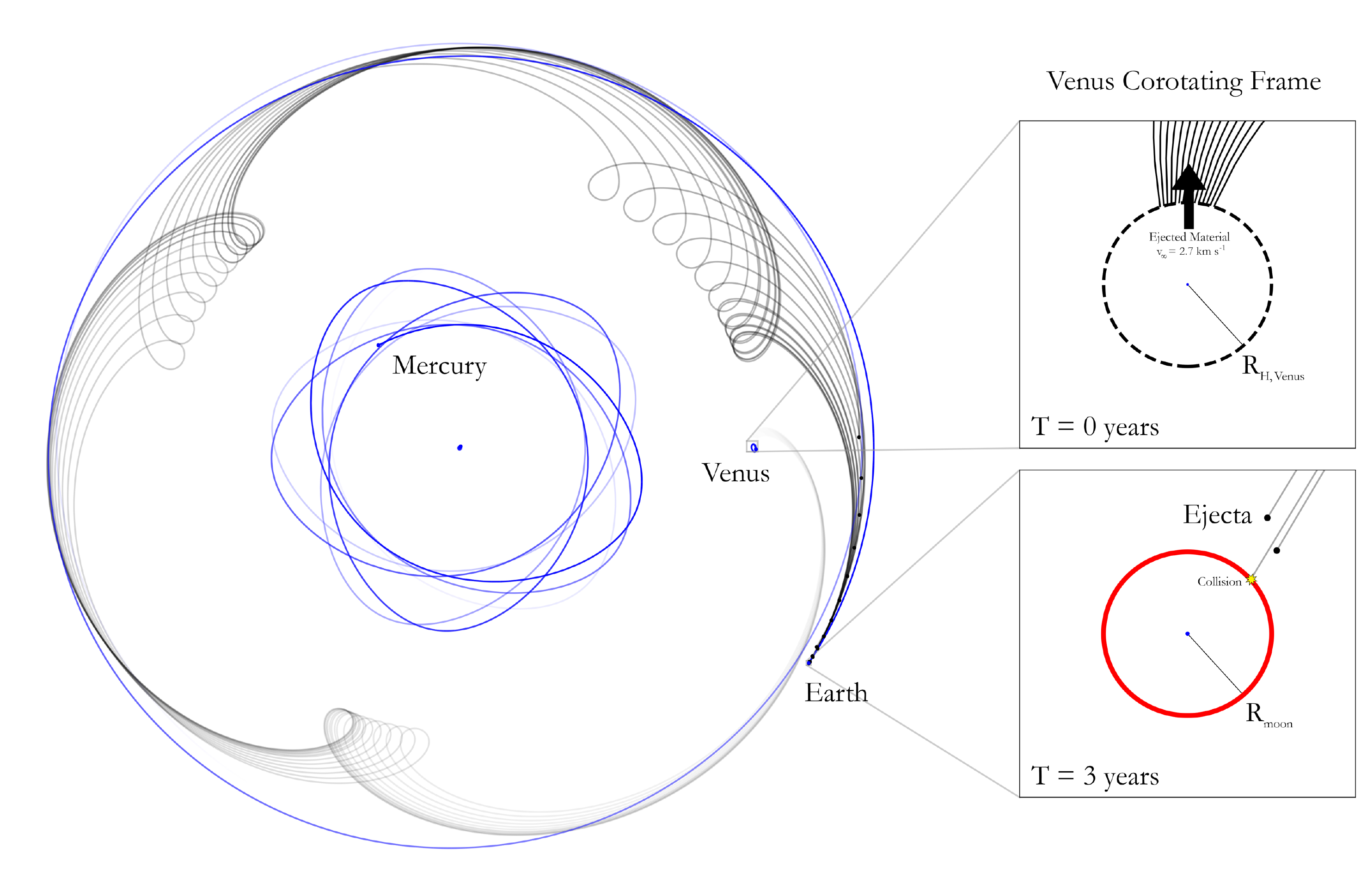}
    \caption{Orbital trajectories of ejecta launched radially from the leading surface of Venus. Ejecta are launched from the Hill Sphere with residual velocities of $2.7$ \kms\ (top zoom-in panel). These conditions give ejecta an elliptical orbit with apocenter at approximately 1 AU, thus crossing Earth's orbit. The residual velocity is low enough such that the fragments may survive collision with the Moon (bottom zoom-in panel). Trajectories are depicted in the Venus corotating frame. Planetary orbits are depicted in blue, while ejecta orbits are in black. Ejecta bulk circular motion result from the difference of their mean motions with that of Venus. The smaller loops arise since the ejecta orbits are highly eccentric. The simulation was integrated for three years.}
    \label{fig:tr}
\end{figure*}

{Ejecta} may melt or fragment when they impact the Lunar surface. Unlike Earth, the Moon lacks an atmosphere to slow down high-velocity projectiles. However, it has a much lower escape velocity and minimum impact speed of about 2.4 \kms, {as well as a more porous surface}. Several studies investigate the aftermath of an impact on the Moon, mainly in the context of Earth-origin fragments ejected during LHB, and potential of recovering information about the early state of Earth \citep{Armstrong2002, Crawford2008, Bland2008, Halim2020}. Oblique ($\lesssim 45^{\circ}$) impacts at speeds $\lesssim7$ \kms\ avoid melting in nearly all cases \citep{Bland2008}. While the leading edge of a $5$ \kms, face-on {projectile} can experience nearly 20 GPa of pressure, the trailing edge almost always experiences less than 10 GPa \citep{Crawford2008}. In our impact scenario, any ejecta with $v_{\infty}$ near $2.7$ \kms\ plus the heliocentric orbital velocity of Venus undergo precise transfer to Earth's orbit (Figure~\ref{fig:tr}). The velocity at apocenter is $27.3$ \kms, and the heliocentric orbital velocity of Earth is 29.78 \kms. Taking the difference of these velocities, and summing it in quadrature with the Moon's escape velocity indicates a significant portion of material will impact the Moon at $\lesssim5$ \kms. The literature suggests most of this material should survive impact. We investigated the distribution of velocities with which ejecta enter the Earth-Moon system through Simulation 3. Collisions with ejecta occur with 16$^{\rm th}$, 50$^{\rm th}$ and 84$^{\rm th}$ percentiles of 3.5, 6.3 and 11.2 \kms\ speed relative to Earth. The collision speed with the Moon involves the addition of the Moon's relative orbital speed (up to $\pm 1$ \kms), plus its escape velocity added in quadrature.

Over time, these ejecta will be become buried within the Lunar regolith, the Moon's unconsolidated, fragmented surface layer. The regolith evolves {through several processes, including:  cratering, melting, fracturing of surface rocks from projectile bombardment \citep{McKay1991}, and secondary impacts \citep{Costello2018}. There are surface interactions with cosmic-rays and the solar wind \citep{McKay1991}, as well as the dirunal temperature cycle \citep{Vasavada2012}.} The largest and earliest rock fragments may lie in a zone beneath the regolith, known as the megaregolith. \citet{McKay2010} notes that this region {could be} rich in information regarding the early solar system, and could {even} contain pieces of meteorites from Venus. In any case, a search for Venusian meteorites likely requires several meters of excavation, and screening of the largest rocks in the regolith. For Earth fragment recovery, \citet{Armstrong2002} recommend at least 300 m of excavation, or starting in $\gtrsim 3$ km diameter craters. \citet{Richardson2020} present 1 km as a lower-bound for the depth of the upper megaregolith. {We adopt this value for our abundance estimates of 0.3 ppm and 0.1 ppm} for Venusian and Earth material respectively. {Accessing this layer would be far more challenging than surveying the uppermost regolith.}

\subsection{Identification and Constraints on Transfer}
\label{sec:id}

Isotopic analyses can eliminate Lunar, Earth and Martian origins of a meteorite displaying terrestrial-like petrology \citep[e.g.][]{Papike2003, Karner2006, Joy2016}. Excess meteorite concentrations may be found on the leading side and poles of the Moon \citep{Armstrong2010}. Venusian material may be located within brecciated Lunar samples, which contain melt and debris from impacts. The projectile may have been the ejecta particle itself, or another meteorite which upturned the deep Lunar regolith. Venusian meteorites would be achondrites, lacking the millimeter-sized mineral pockets (chondrules) and solar composition often found in chondrites \citep{Weisberg2006}. An oxygen fractionation isotope analysis \citep{Clayton1996, Clayton2003} is prudent for identifying Venusian meteorites, and characterizing ancient Venus. \citet{Greenwood2017} present a comprehensive overview of technology needed for testing, as well as physical properties that may be constrained by the measurements. Namely, mass-dependent oxygen fractionation traces chemical and geological processes, such that $^{18}$O/$^{16}$O and $^{17}$O/$^{16}$O vary with a slope of $\sim 0.5$. Mass-independent fractionation is less understood, and traces primordial processes involving water ice and silicates, CO self shielding, accretion and differentiation, and abundance variations in the solar nebula or giant molecular cloud \citep[][and references therein]{Greenwood2017}. Nevertheless, mass-independent fractionation generates oxygen isotope ratios inherent to the origin body of a meteorite. Joint measurements of $\delta^{18}$O and $\delta^{17}$O unambiguously separate SNC (Martian) and HED (asteroid) meteorites from Earth and Lunar rocks \citep{Clayton2003}. Laser fluorination with an infrared CO$_2$ laser is the predominant technique for measuring oxygen fractionation \citep[][]{Miller1999, Franchi1999, Greenwood2014, Starkey2016}. Alternatively, one may use secondary ion mass spectrometry, which is generally of lower measurement precision. Also, UV laser ablation offers enhanced spatial precision; however, bulk measurements $\delta^{18}$O and $\delta^{17}$O are sufficient to determine the meteorite origin. At least, this analysis would rule out a {Terrestrial, Martian, Lunar or asteroidal origin.} Recovery of multiple Venusian meteorites would eventually point toward a distinct, differentiated body. {While oxygen fractionation is a common go-to analysis, there are other chemical tests that assist with meteorite identification. \citet{Papike2003} show that Fe/Mn ratios in pyroxene and olivine are clearly distinguishable for differentiated bodies, as is percent anorthite in plagioclase feldspar. Other important tracers include Ti, Ni, Co, Cr, and V \citep{Karner2003}. \citet{Joy2016} review some additional techniques that can be used to constrain planetary origin.}

While the bulk of a clast may be analyzed, an emphasis should be placed on zircon grains, which are particularly resilient to weathering, transport and metamorphism \citep{Valley2002}. They offer excellent preservation of the U-Pb crystallization age and $\delta^{18}$O value \citep{Valley2003}. For example, high $\delta^{18}$O in the oldest Earth zircons suggest early presence liquid water on Earth \citep{Mojzsis2001}. \citet{2019Bellucci} use a secondary ion mass spectrometer to measure Ti and rare earth element concentrations in their Lunar clast zircon. Their Ti measurement constrains formation pressures to $6.9\pm1.2$ kbar, and their Ce isotope measurements constrain oxygen fugacity in the source magma. These findings, along with a high crystallization temperature \citep{Ferry2007}, are consistent with an Earth origin. {However, this inference was recently disputed by \citet{Warren2020}. In particular, trace metals (Zn, Ga and Ge) are depleted below expectations for Terrestrial rocks. The abundance ratio Lu/Sm and high abundances of both Ta and Ba are much more consistent with a Lunar origin.}

Following an oxygen fractionation analysis, several other chemical and isotopic analyses would provide important information on a Venusian meteorite, as they have for achronditic meteorites arriving from the Moon, Mars and Vesta. For example, the Martian meteorite ALH84001 gained notoriety for claims of potential biosignatures \citep{McKay1996}. Ratios of Sm-Nd and Rb-Sr reveal when ALH84001 cooled and crystallized \citep{Nyquist1995, Jagoutz1994}. The K-Ar ratio dates shock events \citep{Treiman1995}, such as the impact event that may have launched ALH84001. Isotopes of He, Ne, Ar, Kr, and Be place the Cosmic Ray Exposure (CRE) age at around 15 Myr \citep{Eugster2006}. It was deemed of Martian origin based on petrographic similarities to other SNC meteorites \citep{Mittlefehldt1994}. Achondrites also carry the composition of their parent's atmosphere. For example, abundances of Ar and Xe isotopes in EET79001 match those in the Martian atmosphere, as established by the Viking rover \citep{Bogard1983b}. The origins of some meteorites remain more ambiguous. \citet{Irving2013} found the bulk composition of NWA 7325 consistent with Mercury, based on measurements from MESSENGER. However, abundances of Al and Pb-Mg in NWA 7325 suggest a much older age, and \citet{Barrat2015} argue it was more likely ejected from an early Solar System planetesimal. {Additional studies such as \citet{Weber2016} and \citet{Koefoed2016} find evidence against Mercurian origin, but agree on its unusual nature}. The rare class of Urelite meteorites (comprised of carbon-rich olivine or pyroxenes) also lack a well-constrained source \citep{Berkley1980}. Nevertheless, the analyses above give some basis for helping identify a Venusian sample.  A Venusian meteorite would likely have CRE age $>1$ Myr. Through our $N$-body simulations, we confirm that the vast majority of Earth ejecta which returns to Earth does so within $\sim$0.5 Myr \citet{Worth2013}. Ejecta from Venus generally take more time to reach Earth (only about a third arrive in the first million years after ejection); impacts with the Moon should also be subject to these timescales. \citet{Jourdan2017} note that Venusian meteorites would have unique Ar and U‐Th ages, but this applies to rocks that may have been ejected from modern Venus (after its resurfacing event). 

The ratio of median abundances of lightly-shocked Venusian and Earth material on the Moon, as estimated by our Monte-Carlo analysis, is approximately $3-4$. {While \citet{Warren2020} indicate the zircon analyzed by \citet{2019Bellucci} is most likely of Lunar origin, in the event that it did not originate on the Moon, our analysis indicates Venus is a likely alternate point of origin. An oxygen isotope ratio analysis would test this hypothesis.} Progressively building a library of non-Lunar rocks found in the Lunar regolith will constrain the ratio of Venusian to Earth content, and in turn, Venus' water history. A ratio of near $4$ holds if Venus lost its water content in the past 3.5 Gyr. If Venus lost its water $\sim 4$ Gya, then we would expect a couple Venusian samples for every identified Earth sample. A lack of Venusian rocks in a large collection of hundreds of Earth rocks points toward very early water loss $\gtrsim 4.4$ Gya, or falsification of our proposed model of transport. One key assumption of our model is that Earth and Venus experience comparable fluxes of asteroid impacts; however, an updated value would probably not change the ratio dramatically. We have neglected possible contributions of comet impacts, which would likely increase the Earth abundance on the Moon if they do contribute to spallation. Also, the asteroid size and impact speed distributions we have assumed are valid only for recent times. A more likely source of falsification is that spall ejection from Earth and Venus is, in reality, very limited. We assume Equation~\ref{eqn:frac} for estimating the quantity of such material. However, the range of pressures and velocities involved in a large asteroid impact far exceed those intended by the hydrodynamical model \citep{Melosh1984}; spallation would still hold true for the smaller-scale impacts that occur on Mars. If this is indeed the case, then nearly all non-Lunar material on the Moon will be heavily-shocked Earth material.

\section{Conclusion}

With renewed Lunar exploration on the near horizon, we posit that meteorite acquisition and identification will answer an important outstanding question about the history of Venus. We examine the ejection of material during an asteroid impact through well-established hydrodynamical prescriptions. Our state-of-the-art $N$-body simulations predict $\gtrsim0.07\:\%$ of material ejected from the surface of Venus will reach the Moon's surface. This small fraction is augmented by immense amounts of ejected material (of order $10^{13}$ kg) escaping a thin, Earth-like atmosphere. 

Three important processes work in favor of recovering Venusian meteorites. First, much of the ejected material is minimally shocked, due to shock-wave interference and spallation. Second, meteorite fragments are likely to survive their impact on the Moon, given their relatively low impact-velocities. They are even more likely to survive oblique impacts. Finally, the Moon's regolith is relatively shallow. It is amenable to excavation, particularly in existing craters; although excavation of the deeper megaregolith presents a more challenging scenario. The Earth on the other hand would have rapidly destroyed any ancient Venusian meteorites. Our findings indicate that {\it in situ} analysis or sample return missions, with particular focus on zircon-grains and oxygen isotope fractionation, have a high potential of identifying ancient Venusian meteorites. 
\\
\\
\\
{We thank the anonymous referees for their constructive comments, which helped improve the clarity of the manuscript. We would like to acknowledge Jay Ague for insightful discussions on petrology and isotopic dating, and also for reviewing the text. We also thank Tolman Geffs for a careful review and helpful discussions.}

\bibliographystyle{aasjournal} 
\bibliography{main} 

\end{document}